\documentclass[twocolumn,superscriptaddress,prb]{revtex4-2}

\usepackage{graphicx}
\usepackage{amssymb}
\usepackage{amsmath}
\usepackage{hyperref}
\usepackage{ragged2e} 
\usepackage{placeins}
\usepackage{array} 
\usepackage{listings} 
\usepackage{multirow} 
\usepackage[labelformat=simple]{subcaption}
\usepackage{xcolor}
\DeclareCaptionJustification{justified}{\justifying}
\newcommand{\ket}[1]{\lvert #1 \rangle}
\newcommand{\ndata}{n_{\mathrm{d}}}
\newcommand{\nanc}{n_{\mathrm{a}}}
\newcommand{\GF}{\mathrm{GF}(2)}
\newcommand{\wtS}{\mathrm{wt}_{\mathcal{S}}}

\begin{document}
\title{Agentic AI for Discovering Fault-Tolerant Quantum Circuits}
\author{Aritra Lahiri}
\email[]{aritra.lahiri@mpl.mpg.de}
\affiliation{Max Planck Institute for the Science of Light, 91058 Erlangen, Germany}
\author{Florian Marquardt}
\email[]{florian.marquardt@mpl.mpg.de}
\affiliation{Max Planck Institute for the Science of Light, 91058 Erlangen, Germany}
\affiliation{Department of Physics, Friedrich-Alexander University Erlangen-N\"urnberg, 91058 Erlangen, Germany}
\date{\today}

\begin{abstract}
{
Scaling quantum computing toward practical quantum advantage depends critically on the 
efficiency of circuits that operate on information encoded by quantum error-correcting codes. 
Yet, these circuits are largely designed by hand for individual cases, with some recent works 
incorporating machine learning for certain aspects. We formulate this problem as agentic 
evolutionary program synthesis, wherein a large language model, leveraging its pretraining 
knowledge and tool-use proficiency, evolves programs that define Clifford circuits. Crucially, 
the agent tests and refines its proposals by freely running Python code of its own, alongside 
a verifier that exhaustively checks fault tolerance. Our approach consistently finds 
fault-tolerant circuits for a wide range of curated tasks spanning the primary requirements 
of a quantum computer: logical state preparation, logical gate design, and logical 
routines such as code-switching and code-growth. The resulting programs are interpretable, 
exposing the design logic behind each circuit.
}
\end{abstract}
\maketitle

\section{Introduction}
\label{sec:intro}

Quantum error correction (QEC) codes protect quantum systems and algorithms from noise 
and decoherence by encoding quantum information in redundant and entangled noisy physical 
qubits~\cite{Gottesman2010}. While recent experiments have demonstrated key ingredients of 
universal fault-tolerant quantum computing~\cite{Acharya2025,Bluvstein2024,Bluvstein2025,
GoogleQAI2025Cultivation}, scaling them to the circuit sizes required for practical quantum 
advantage will require substantial advances in fault-tolerant circuit design. This typically
involves a stack of three tasks, in increasing level of complexity: 
preparing logical states, implementing logical gates, and composing them to realize some 
logical routine. For each of these tasks, fault tolerance is a necessity, as a single fault 
anywhere in these circuits can spread into an error larger than what the code can correct, 
causing a logical error. Codes of larger distance tolerate more faults, but their circuits 
require more gates, which increases the possibility of faults. Designing a circuit therefore 
entails simultaneous control over the circuit's objective and fault tolerance. These design 
problems have so far been solved mostly by hand, separately for each task, QEC code, etc., 
which rapidly becomes intractable for large circuits. Automated discovery has, quite recently, 
started making progress. Satisfiability solving yields gate- or depth-optimal circuits 
preparing logical states of small CSS codes together with the verification circuits for 
fault tolerance~\cite{Peham2025}. Reinforcement learning (RL) approaches train a neural
network to build circuits by placing gates, or suitably transforming them, getting rewarded 
as the circuit approaches its target~\cite{Kremer2024,Rietsch2024,Doherty2026,Olle2024,
Fosel2018,Fosel2021,Ruiz2025}. Reference~\cite{Zen2025} extends it to fault-tolerant logical 
state preparation using flag qubits. So far, these methods have been limited to distance-five
codes, and have not been applied to logical gates or higher-level routines. Additionally, 
RL requires orders of magnitude more circuit simulations than our large-language-model 
(LLM) based method introduced below. Next up on the stack, logical gates constitute a 
seemingly harder problem as the gate circuit must map each codeword to its image under the 
desired gate operation. Existing works~\cite{Rengaswamy2020,Kuehnke2025,Chen2022,Meyer2026,
Sayginel2025,Benhemou2026} have been mostly limited to small codes, with fault tolerance 
either imposed under restrictions of transversality, or added separately afterwards, if at 
all. Finally, at the last level of the stack, these two fundamental tasks, logical state 
preparation and logical gate design, must be orchestrated to realize logical routines. 
At the level of physical qubits, there are only limited studies in this direction~\cite{Shutty2022}. 

\begin{figure}[t]
\centering
\includegraphics[width=\columnwidth]{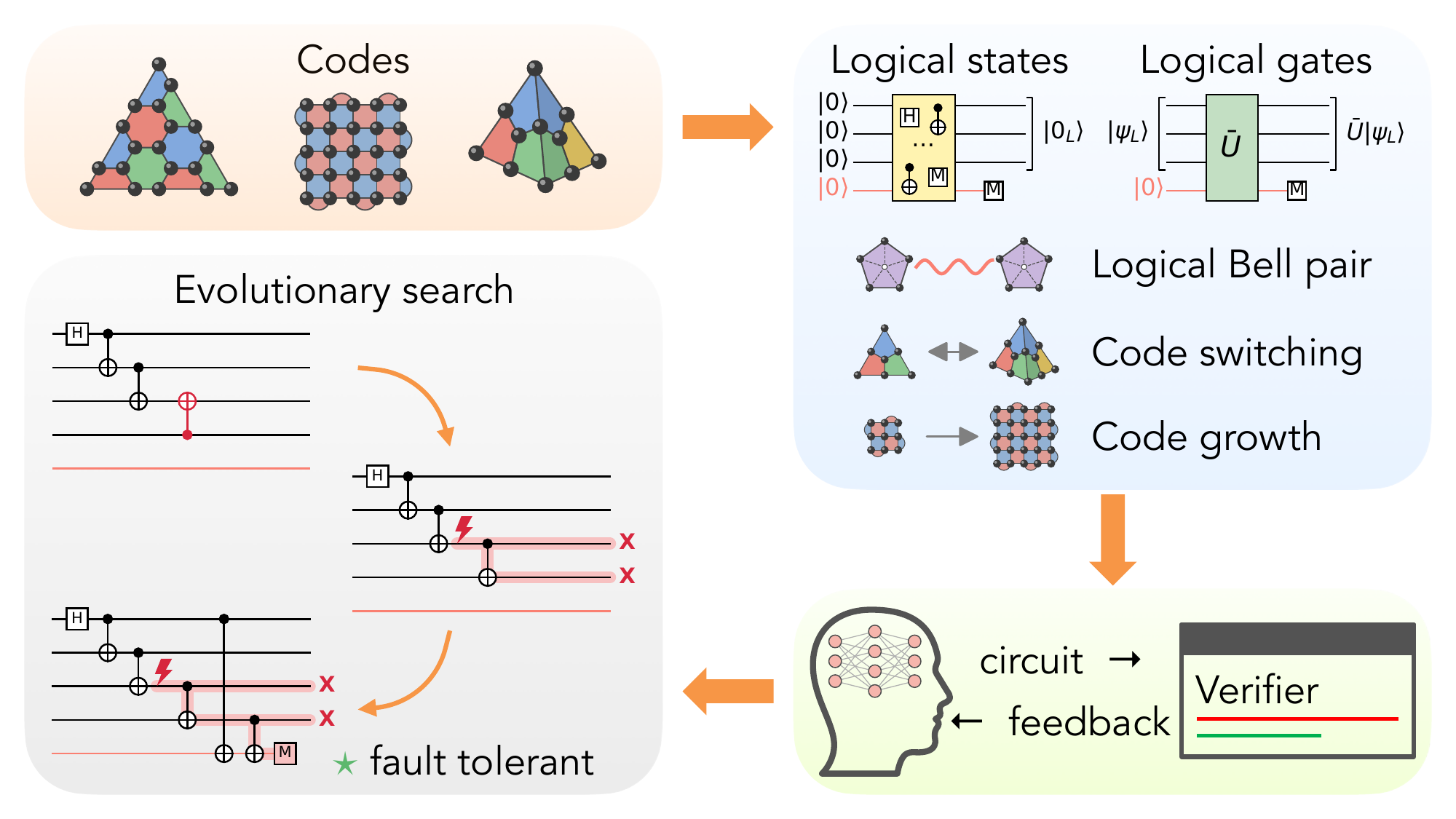}
\caption{Starting with some stabilizer codes (top left), followed by a task specification 
(top right) that includes logical state preparation, logical gate design, or further 
applications such as code-switching, etc., a large language model (LLM) discovers 
fault-tolerant circuits. The LLM is a part of an agentic loop (bottom right), with
access to a circuit verification tool to check for fault tolerance. This is incorporated 
in an evolutionary search (bottom left), where the LLM's proposals are refined iteratively.}
\label{fig:llm_circ}
\end{figure}

To this end, we harness the capabilities of modern frontier LLMs for automated generation of 
fault-tolerant circuits at the level of physical qubits for this entire stack of tasks, logical 
state preparation, logical gate design, as well as logical routines such as code-switching,
code-growth, etc., while offering the tunability for any code and connectivity. LLMs have 
already been applied to quantum circuit generation, with mixed success, either following 
supervised~\cite{Dupuis2024} or RL-based fine-tuning~\cite{Dupuis2025}, or without any 
domain-specific fine-tuning~\cite{Guo2025,CruzBenito2026,Paz2026}. Notably, they have so 
far not been successful at fault tolerance. All of these primarily use the LLM for one-shot 
generation. Outside quantum computing, a different paradigm was introduced with 
FunSearch~\cite{RomeraParedes2024}, and its successor AlphaEvolve~\cite{Novikov2025}, 
which use an LLM as a mutation operator over functions/programs, coupled to an evaluator, 
iteratively refining programs in an evolutionary search. They have produced advances 
over human-designed benchmarks across a range of problems. More recently, 
Refs.~\cite{LiuMarquardt2026,CruzBenito2026b} have used this approach for the discovery 
of quantum low-density parity-check codes. We use this as a starting point, developing 
an agentic FunSearch approach to search over programs representing circuits. Our 
approach rests on the premise that modern frontier LLMs, armed with suitable tools, 
already possess the necessary competence for generic physics tasks~\cite{Nagele2026,Pan2025}.
As depicted in Fig.~\ref{fig:llm_circ}, starting with the physical setting (QEC code, 
qubit layout) and the task specification in the prompt, an LLM generates quantum circuits within an 
evolutionary search using a verification tool to check for circuit validity and fault 
tolerance. Each circuit is specified as a Python function through a simple, custom-made API. 
We focus on dynamic Clifford circuits, which are Clifford circuits with mid-circuit 
measurements and Pauli feedforward gates conditioned on their outcomes. This lets the 
LLM discover the corrections when faults are detected instead of discarding the faulty 
state. This is particularly crucial for tasks such as logical gates, where discarding 
a shot from a gate midway through a long protocol would force one to restart the entire 
protocol from scratch.

The remainder of this manuscript is organized as follows: Sec.~\ref{sec:funsearch} 
introduces our agentic FunSearch and its key components: the prompt, which specifies 
the task, qubit layout, gates, etc.; the verifier, which checks the circuit's validity
and quantifies fault tolerance; and the evolutionary search. Secs.~\ref{sec:states}, 
\ref{sec:gates}, and~\ref{sec:applications} present the results for logical state 
preparation, logical gates, and some applications involving higher-level logical routines 
such as code-switching and code-growth, respectively. We discuss some key features of our 
agentic FunSearch in Sec.~\ref{sec:disc}, and then conclude with an outlook in 
Sec.~\ref{sec:conc}.

\section{Agentic FunSearch for Quantum Circuits}
\label{sec:funsearch}

\begin{figure*}[t]
\centering
\includegraphics[width=\textwidth]{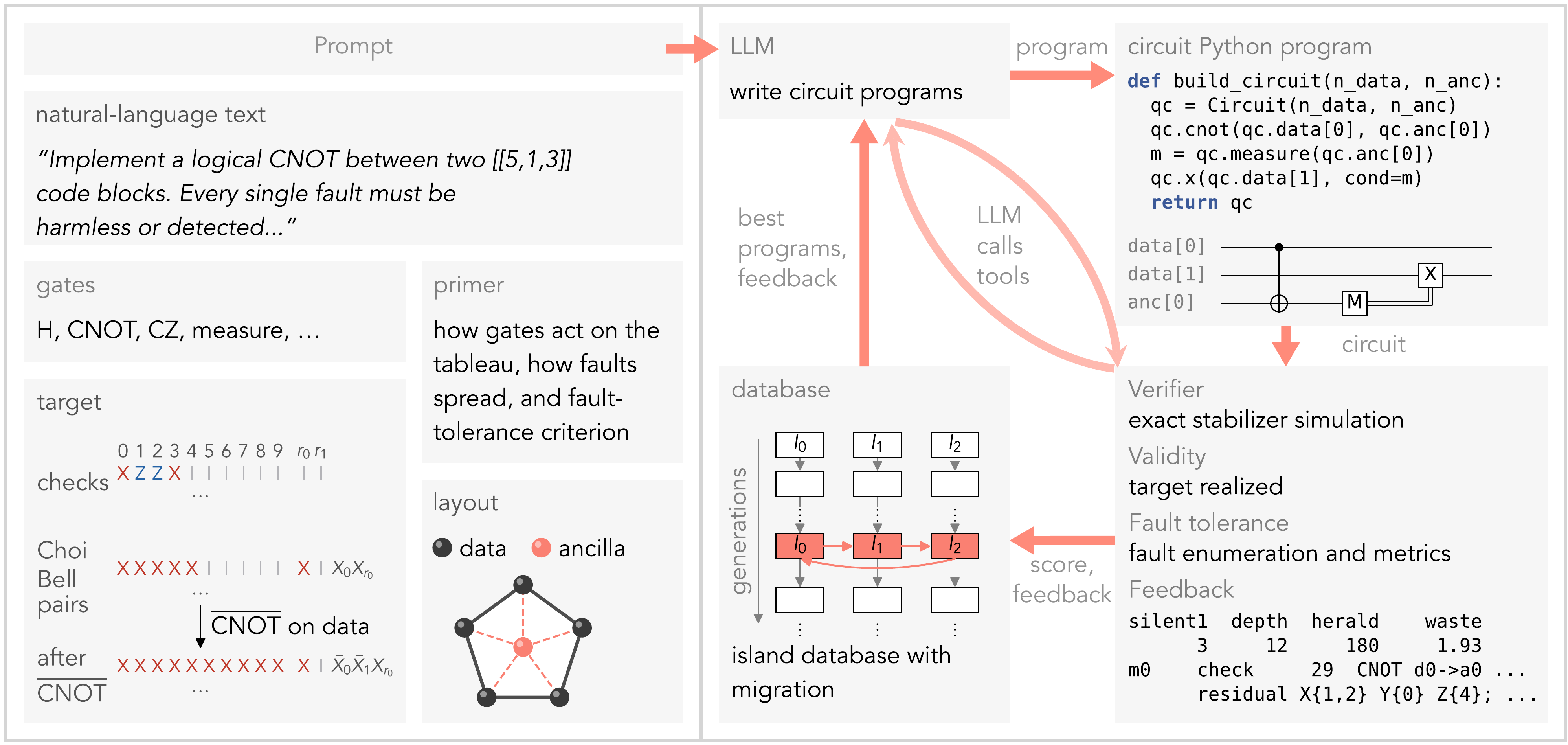}
\caption{Overview of the agentic FunSearch method applied to quantum circuits. Left: the prompt, 
illustrated for a logical CNOT ($\overline{\mathrm{CNOT}}$) between two logical qubits encoded 
by the $[[5,1,3]]$ perfect code (see Sec.~\ref{sec:gates}). It starts with a natural-language 
description of the task, followed by a list of the target stabilizers, here the Choi state 
stabilizers on all physical and reference qubits. These include the code stabilizers, labeled 
``checks'', and the stabilizers of the Bell pair between the logical and the reference qubits 
after applying the desired logical operation (here, $\overline{\mathrm{CNOT}}$), labeled ``after 
$\overline{\mathrm{CNOT}}$". Additionally, the prompt also specifies the qubit layout, including 
the locations of the data and ancilla qubits, the allowed couplings for two-qubit gates, the gate 
set, and a primer on the relevant physics and the fault tolerance requirement. Right: the agentic 
FunSearch loop. The LLM writes a Python program defining a circuit following our custom API. 
The verifier checks if the circuit achieves the target, and obtains the fault tolerance metrics. 
These metrics, along with the packed (maximizing parallel qubit utilization) circuit, are attached 
to each program as feedback. The program is then added to the island database, and ranked by these 
metrics (see Eqs.~\eqref{eq:score} and~\eqref{eq:score_gates}). The best programs from the islands, 
along with their feedback, are included in the prompt to the LLM in the subsequent generation. In 
each generation, we instruct the LLM to write scratchpad notes, which serve as memory and are 
appended to the next generation's prompt. The best programs from each island migrate to the 
neighbouring islands periodically. The LLM also calls the verifier directly through tool calls, 
which lets it test its proposals before submitting its program in each generation. After some 
generations, we select the best program as the solution.}
\label{fig:funsearch_qc}
\end{figure*}

In this section, we describe our ``agentic FunSearch'' for the automated design of quantum 
circuits. As in the original FunSearch, we perform an evolutionary search over programs, 
in which a pretrained LLM acts as the mutation operator and a verifier scores each program~\cite{RomeraParedes2024}.
In our case, each program is a Python function representing a 
quantum circuit, and the verifier checks if the circuit reaches its target and quantifies 
its fault tolerance. We transcend the original FunSearch by harnessing the agentic capabilities 
of frontier LLMs. Rather than merely returning a new program, the LLM can call the verifier 
as a tool, so that it may test and amend its proposals before submitting its solution. 
Our implementation, illustrated in Fig.~\ref{fig:funsearch_qc}, consists of three parts: 
the LLM writes a program defining a circuit following the task specification in its prompt 
(employing the agentic loop just mentioned), a verifier checks the validity of the circuit 
and provides metrics characterizing the fault tolerance of the circuit, and a database 
stores the best programs and feeds them back to the LLM. We introduce each part below, 
and then explain the agentic FunSearch loop. We provide a concise pseudocode of our 
workflow in Appendix~\ref{app:pseudo}.

The prompt provides the physical setting of the problem, a detailed specification of the 
task and circuit constraints, and a description of what the LLM must generate. We focus on 
stabilizer circuits~\cite{Gottesman2010,AaronsonGottesman2004}, so the task is presented in 
the stabilizer formalism. An $n$-qubit stabilizer state is the unique joint $+1$ eigenstate 
of $n$ independent, commuting Pauli operators, known as stabilizers. Each stabilizer is 
stored as a sign and two binary vectors $x,z\in\GF^n$ that denote the qubits on which 
it acts with $X$ or $Z$, respectively. These $n$ stabilizers are thus represented as an
$n\times (2n+1)$ stabilizer tableau. As the stabilizers of a state are not unique, we use the 
canonical tableau. A Clifford gate updates the tableau by binary operations on its columns 
and signs. The target is a stabilizer state $\ket{\psi_\star}$, whose stabilizers are listed 
in the prompt. We primarily focus on targets encoded in a QEC code, denoted $[[n,k,d]]$, 
with $n$ physical qubits, $k$ logical qubits and distance $d$. We also define 
$t=\lfloor (d-1)/2\rfloor$ as the maximum weight of a Pauli error that the code can correct. 
For logical state preparation, the target stabilizers are the $n-k$ code stabilizers and the 
$k$ logical operators that fix the state, e.g., the logical $Z$, $\bar Z$, for $\ket{0_L}$, 
where the bar denotes a logical operator. These are provided by the user as part of the prompt 
to the LLM. For designing a logical gate, $\bar U$, we must ensure that it implements the 
desired logical operation on every codeword. While this is a seemingly harder task than 
logical state preparation, the channel-state duality~\cite{Jamiolkowski1972,Choi1975} 
reduces this to a state preparation problem. Specifically, we introduce reference qubits 
such that each logical qubit $j$ is coupled with a reference qubit $r_j$ in a Bell state. 
The target becomes the logical Choi state of 
$\bar U$~\cite{NielsenChuang1997,GottesmanChuang1999,Peham2023}, generated by the code 
stabilizers, along with the stabilizers of the Bell pair between the logical qubits and 
the reference qubits, $\bar U\bar X_j\bar U^\dagger\otimes X_{r_j}$ and 
$\bar U\bar Z_j\bar U^\dagger\otimes Z_{r_j}$ (Appendix~\ref{app:gatechoi}). In this 
case the code stabilizers, all the logical operators $\bar X_j$ and $\bar Z_j$, and 
$\bar U$ are provided by the user. Our code obtains the Bell-pair stabilizers 
mentioned above, and appends them to the prompt. Besides the target, the 
prompt specifies the qubit layout, i.e., the positions of the data and ancilla qubits, 
and the qubit pairs on which two-qubit gates are permitted. It also specifies the gate 
set, a short primer on how gates and measurements alter the stabilizer tableau, and 
states the fault tolerance requirement as described below.

The evolved artifact is a Python function that represents a circuit~\cite{Sarra2024}. It takes 
the number of data and ancilla qubits, $(\ndata,\nanc)$, as arguments instead of hardcoding them, 
so that one program can describe a whole family of circuits wherever applicable, e.g., the surface
code at several distances. Programs also express conditional logic naturally, which lets them 
represent dynamic Clifford circuits. Finally, since LLMs are trained extensively on code, 
they are well suited to writing programs. Moreover, unlike a list of gates, a program is 
interpretable~\cite{RomeraParedes2024}, and the prompt also instructs the LLM to explain the 
design logic in the docstring (see Appendices~\ref{app:prompt} and~\ref{app:llmprogram}). The 
program builds the circuit by calling a small custom-made circuit class, whose API is described 
to the LLM in its prompt. This class is initialized with the qubit layout chosen by us. This 
is accessible to the program via the API. We describe the layouts as they arise in later sections. 
The circuit class provides the Clifford gates, $Z$-basis measurement of ancillae, as well as 
conditional $X$ or $Z$ gates where the condition can be the parity/XOR of any number of 
measurement outcomes. It records these operations as a circuit when the Python function is 
executed, and the verifier uses it to verify the circuit. This involves two steps: First, 
it checks that the circuit is valid, i.e., the circuit leaves the data qubits in the target 
state (logical Choi state in the case of a gate, and other applications such as code-switching), 
and disentangled from the ancillae, for every combination of random measurement 
outcomes~\cite{Kliuchnikov2023}. Second, the verifier checks for fault tolerance. We implement 
flag fault tolerance, wherein some ancilla qubits serve as flag qubits which detect 
faults~\cite{ChaoReichardt2018,ChamberlandBeverland2018}. Specifically, the ancillae whose 
measurements are deterministic serve as flags, or detectors, as their outcomes are fixed in 
the fault-free circuit, so any deviation signals a fault. This also includes fixed XORs of 
random outcomes, e.g., a repeated measurement of an ancilla whose first outcome was random. 
Note that the faults cannot change whether a measurement is deterministic or random. This 
is because this classification depends only on the symplectic ($X$ and $Z$) part of the 
stabilizer tableau, specifically, a measurement is deterministic if the measured operator 
belongs to the stabilizer group up to some signs, and random otherwise. Since the faults 
are modeled as Paulis inserted after the failing gates (discussed below), which do not 
alter the symplectic part of the tableau, they cannot change the classification of the 
measurements. We briefly describe the fault tolerance criterion below, and provide additional 
details in Appendix~\ref{app:faulttol}.

The verifier enumerates the faults of a valid circuit, similar to the standard circuit-level 
depolarizing noise model, accounting for gate failures, input errors, and measurement 
errors, each happening with a physical error rate $\sim p$. A fault is a single non-identity 
Pauli inserted after a non-Pauli gate or on the input qubits, or a flipped measurement 
outcome. Pauli gates are omitted as they are typically tracked in software~\cite{Knill2005}. 
We call $i$ simultaneous faults an event of order $i$, which occurs with probability $\sim p^i$. 
A single fault is a first-order event, and events of order $i\geq 2$ are higher-order events. 
For a code of distance $d$, we only enumerate events of order $i\leq t$ (for the even-distance 
qLDPC codes, up to $d/2$). Events of order greater than $t$ occur with probability 
$\lesssim p^{t+1}$, and therefore they do not alter the exponent of the logical error rate 
of the code, $p_L\sim p^{t+1}$. Each fault is propagated to the end of the circuit, leaving 
behind a residual Pauli $E$ on the data qubits, and flipping some detectors, and we call 
this set of flipped detectors its pattern. The residual and pattern of a higher-order 
event are the product of its faults' residuals, and the XOR of their patterns, 
respectively (Appendix~\ref{app:faulttol}). Two residuals, possibly with different weights, 
act identically on the target if they differ by a Pauli $G$ on the data qubits that 
stabilizes the target. These Paulis form a group $\mathcal{G}$. Hence, a residual is judged 
by its reduced weight $\mathrm{wt}_{\mathcal{G}}(E)=\min_{G\in \mathcal{G}} \mathrm{wt}(EG)$, 
which only depends on how the residual acts on the target. For logical state preparation, 
$\mathcal{G}$ is the stabilizer group of the target state, which is generated by the $n-k$ 
code checks and $k$ logical operators that together identify the target state. For logical 
gate design, code-switching, code-growth, etc., $\mathcal{G}$ is the group generated by the 
output code checks alone, as these circuits must implement a fixed logical operation 
on the input logical state and any logical operator acting on a residual would alter that 
operation for some inputs. The circuit is fault-tolerant when every event of order $i\leq t$
either fires a detector (logical gates have a different criterion, as discussed below) or 
leaves a residual with~\cite{AGP2006,Gottesman2010,ChamberlandBeverland2018}
\begin{equation}
  \mathrm{wt}_{\mathcal{G}}(E)\;\le\; i .
  \label{eq:criterion}
\end{equation}
A residual which exceeds this bound is denoted ``heavy'', and ``light'' otherwise.
We note that for $\ket{0_L}$ ($\ket{+_L}$) of a CSS code, only the $X$ ($Z$) part of 
each residual needs to satisfy the bound in Eq.~\eqref{eq:criterion}. For logical gates 
and code-growth on CSS codes, neither part can be ignored, but we check $X$ and $Z$
parts of each residual separately against the bound as they are corrected independently 
(Appendix~\ref{app:faulttol}). Since we consider flag fault tolerance with ancilla 
measurements as fault detectors, each event is judged not only by its residual's weight 
using Eq.~\eqref{eq:criterion}, but also by its detector pattern and, for logical gates, 
the syndrome measurements from the subsequent QEC rounds as well. We adopt two distinct 
ways to classify the events, depending on the task. 

We start with logical state preparation. Since it typically constitutes the starting 
point of QEC protocols, it is affordable to discard a faulty shot and rerun the circuit 
until a shot is accepted. Thus, we simply discard any shot where ancilla measurements 
indicate a heavy event, which implies we only require an event's 
residual weight and its detector pattern to classify it. Accordingly, each event is 
classified as ``correctable'' if its residual satisfies the bound in Eq.~\eqref{eq:criterion}, 
``silent'' if it exceeds the bound but no detector is triggered, ``heralded'' if it exceeds 
the bound but at least one detector is triggered resulting in a non-trivial pattern, and 
``waste'' if it is correctable but it still triggers a pattern that is shared with a 
heralded event, so post-selection discards it as well. The primary goal of our search 
is to drive the number of silent events down to zero as they compromise fault tolerance. 
Idling faults, which are Pauli faults on a qubit while it is waiting between two 
consecutive operations, are not enumerated separately, as every idling fault coincides with 
one of the faults already enumerated by the verifier. They only increase the prefactor 
of the logical error rate~\cite{Gottesman2010}. Since the number of idling faults is 
commensurate with the depth of the circuit, reducing the depth constitutes the secondary 
goal of our search. The heralded and wasted events only impact the acceptance rate, and 
since discarding a shot is acceptable for state preparation, they constitute the 
tertiary and quaternary goals, respectively. Hence, we adopt the following lexicographic 
score to rank the logical state preparation circuits:
\begin{equation}
  \big(\text{silent-cost},\; \text{depth},\; \text{herald},\; \text{waste},\; -\text{match}\big),
  \label{eq:score}
\end{equation}
where $\text{silent-cost}=\sum_{i\le t} \text{silent}_i/(cN)^{i-1}$. 
Here, $\text{silent}_i$ denotes the number of silent events of order $i$, 
$N=\ndata+\nanc$ is the total number of qubits, and $c\sim O(1)$ is an arbitrary constant. 
The factor $(cN)^{i-1}$ compensates for the combinatorial growth in the number of possible 
events of order $i$, preventing higher-order event contributions from dominating the score 
solely due to their multiplicity. The depth is obtained by compressing the circuit to 
maximize parallel utilization of qubits. The calculation of $\text{herald}$ and 
$\text{waste}$ is elaborated in Appendix~\ref{app:faulttol}. They are both obtained only 
for first-order events. This is because we never obtain 
zero heralded and wasted events in practice, and the corresponding counts from higher-order 
events cannot vanish as long as those for the single faults are non-zero.
Moreover, once $\text{silent}_{i=1\ldots t}=0$, all heavy events trigger a pattern, thereby 
ensuring fault tolerance under post-selection. The match is $1$ for every valid circuit and 
otherwise grows with the fraction of measurement branches that reach the target 
(Appendix~\ref{app:frame}), so it matters only for invalid circuits. While this discussion 
was for a search run for a single target, in some cases, we also perform a simultaneous 
search over multiple targets, typically involving the same task but over a family of QEC
codes, e.g., logical state preparation over several code distances for the rotated surface 
code. In this case, a program is valid only if its circuits are valid at every distance. 
Using $S_j$, $D_j$, $H_j$ and $W_j$ to denote the silent-cost, depth, herald and waste 
at the $j$th largest target distance, respectively, the programs are ranked by the 
lexicographic score $(S_1,D_1,H_1,W_1,\allowbreak S_2,S_3,\ldots,\allowbreak D_2,D_3,
\ldots,\allowbreak H_2,H_3,\ldots,\allowbreak W_2,W_3,\ldots)$. This ordering is 
motivated by the fact that the largest distance is the most difficult target, so its full 
score takes precedence. The smaller distances are intended to encourage a size-general 
program, but not to outrank the largest distance.

Now, for logical gates, code-switching, code-growth and other mid-circuit gadgets, which 
reside in the middle of long protocols, discarding a shot would force the entire protocol 
up to that point to repeat. If such a strategy were adopted, in a long protocol with 
several such mid-circuit gadgets, the probability of successfully completing the protocol 
would vanish exponentially with their number~\cite{Reichardt2026,Knill2004,Knill2005}. 
Moreover, a discard caused by an event of order $i$, comprising $i$ simultaneous faults, 
occurs with probability $\sim p^i$, which is parametrically greater than the logical error 
rate $p_L\sim p^{t+1}$ of the QEC code. As such, the discards would become the dominant factor 
restricting the length of the protocol, negating the benefit of the QEC code~\cite{Reichardt2026}. 
Therefore, to overcome this challenge, we adopt a different approach, without post-selection. 
This requires a change in how we classify the events based on their residuals and their 
flag detector patterns. Instead of discarding a shot after detecting the presence of 
heavy events, we correct them. The code check syndrome alone cannot achieve this, as 
a heavy residual can share its syndrome with a light one while differing from it by a 
logical operator. As in flag error correction~\cite{ChamberlandBeverland2018}, the 
detector pattern supplies the missing information~\cite{TansuwannontLeung2022}. 
We therefore assume that in the subsequent QEC round, the decoder uses both the detector 
pattern and the code check syndrome together, henceforth called a bin. If no detector 
is triggered (empty pattern), the decoder decodes the syndrome 
alone as in a standard QEC round, and each event is then judged as silent or correctable, 
using Eq.~\eqref{eq:criterion}. If some detector is triggered, the decoder uses the bin to 
determine a correction. Note that residuals belonging to a bin can come from events of any 
order $i\leq t$. Since they share the same syndrome, they may differ only by an element of 
$\mathcal{G}$ (product of some code checks), or by some logical operator times an element of 
$\mathcal{G}$. If all residuals in this bin differ only by elements of $\mathcal{G}$, they are 
unambiguous and a single correction repairs all of them, including heavy residuals. Otherwise, 
when the residuals of at least a pair of events in the bin differ by some logical operator, the 
bin is ambiguous, as no single correction can correct both. Such an ambiguous bin must contain 
at least one heavy residual. This is because a logical operator $L$ has 
$\mathrm{wt}_{\mathcal{G}}(L)\geq d$, and when a light residual $E$ of an event of order 
$i\leq t$ with $\mathrm{wt}_{\mathcal{G}}(E)\leq i\leq t$ is multiplied by $L$, the product 
must be heavy, $\mathrm{wt}_{\mathcal{G}}(LE)\geq d-t$. Each event in such an ambiguous bin 
must be discarded, hence such a case fails our goal of forgoing post-selection. We call 
such a discarded event ``undecodable'' if its residual is heavy, or ``forfeited'' otherwise. 
Thus, unlike for state preparation, we discard a detected event with a heavy residual not 
based on its pattern, but only when its bin is ambiguous and cannot be corrected. Moreover, 
since every ambiguous bin must contain an undecodable event, having zero undecodable events 
implies zero forfeited events. The bins and their corrections follow directly from our 
verifier's exhaustive check, so each circuit comes with its look-up table for decoding. 
We evaluate both at every order $i\le t$ as described in Appendix~\ref{app:faulttol}, 
with each event weighted by the product of its constituent faults' weights, so that 
$\text{undecodable}_i$ and $\text{forfeited}_i$ are approximately the probabilities, 
in units of $p^i$ at order $i$, of discarding a shot. These are combined into the
weighted sums $\text{undecodable-cost}=\sum_{i\le t}\text{undecodable}_i/(cN)^{i-1}$ and 
$\text{forfeit-cost}$, respectively, in the same way as the silent-cost. We then rank the 
gate circuits by the lexicographic score
\begin{align}
  \big(\text{silent-cost},\; &\text{undecodable-cost},\; \text{forfeit-cost},\nonumber\\
                             &\text{depth},\; -\text{match}\big).\label{eq:score_gates}
\end{align}

\begin{table*}[!htb]
\centering
\setlength{\tabcolsep}{3pt}
\small
\begin{tabular}{cllcccccccc}
\hline\hline
 & code & circuit & connectivity & anc. (used) & det. & silent$_{1/2/\ldots}$ & depth & herald & waste & 2q  \\
\hline
\multirow{2}{*}{\raisebox{-0.4\height}{\includegraphics[scale=0.5]{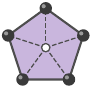}}}
 & $[[5,1,3]]$          & Ref.~\cite{Zen2025}, $\ket{1_L}$  & all-to-all & 2 (2) & 2 & 0 & 12 & 2.47 & 14.67 & \textbf{12}\\
 &                      & \textbf{ours}, $\ket{1_L}$   & all-to-all & 2 (2) & 2 & 0 & \textbf{8} & \textbf{1.73} & \textbf{11.33} & \textbf{12}\\
\hline
\multirow{3}{*}{\raisebox{-0.4\height}{\includegraphics[scale=0.5]{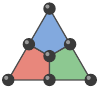}}}
 & $[[7,1,3]]$         & Ref.~\cite{Goto2016}, Fig.~1(c)  & all-to-all & 1 (1) & 1 & 0 & 7  & 2.53 & \textbf{7.00} & \textbf{11}\\
 &                     & \textbf{ours}, multi-size & all-to-all & 3 (3) & 6 & 0 & 7  & \textbf{1.33} & 7.20 & 17\\
 &                     & \textbf{ours}, multi-size$^\dagger$ & all-to-all & 3 (3) & 3 & 0 & \textbf{6} & 1.60 & 10.87 & 15\\
\hline
\multirow{3}{*}{\raisebox{-0.4\height}{\includegraphics[scale=0.35]{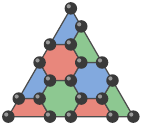}}}
 & $[[19,1,5]]$        & Ref.~\cite{Zen2025}             & all-to-all & 5 (5) & 5 & 0/0 & 16 & 14.67 & 7.13 & 52\\
 &                     & \textbf{ours}, single-size & all-to-all & 4 (4) & 6 & 0/0 & \textbf{8} & \textbf{6.40} & 14.40 & \textbf{50}\\
 &                     & \textbf{ours}, multi-size  & all-to-all & 4 (4) & 10 & 0/0 & 17 & 8.27 & \textbf{0.00} & 77\\
\hline
\multirow{2}{*}{\raisebox{-0.4\height}{\includegraphics[scale=0.25]{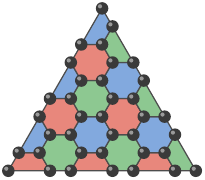}}}
 & $[[37,1,7]]$        & \textbf{ours}, single-size & all-to-all & 7 (7) & 14 & 0/0/0 & \textbf{18} & \textbf{18.40} & \textbf{0.00} & \textbf{148}\\
 &                     & \textbf{ours}, multi-size  & all-to-all & 7 (7) & 25 & 0/0/0 & 23 & 32.00 & 2.53 & 189\\
\hline
\multirow{2}{*}{\raisebox{-0.4\height}{\includegraphics[scale=0.19]{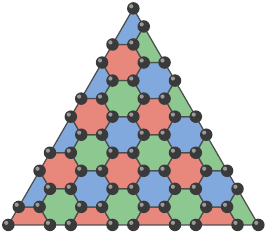}}}
 & $[[61,1,9]]$        & \textbf{ours}, single-size & all-to-all & 11 (11) & 41 & 0/0/0/0 & 32 & \textbf{31.73} & \textbf{0.00} & \textbf{334}\\
 &                     & \textbf{ours}, multi-size  & all-to-all & 11 (11) & 31 & 0/0/0/0 & \textbf{28} & 61.60 & \textbf{0.00} & 361\\
\hline
\multirow{3}{*}{\raisebox{-0.4\height}{\includegraphics[scale=0.5]{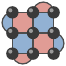}}}
 & $[[9,1,3]]$   & Ref.~\cite{Colmenarez2026}, ME   & native & 8 (4)   & 0  & 16      & 7          & --   & -- & \textbf{12}\\
 &               & \textbf{ours}, multi-size   & native & 8 (8)   & 4  & 0       & \textbf{6} & 1.07 & 9.20 & 27\\
\cline{3-11}
 &               & \textbf{ours}, multi-size   & hex    & 8 (6)   & 2  & 0       & 6          & 3.47 & 5.20 & 18\\
\hline
\multirow{6}{*}{\raisebox{-0.4\height}{\includegraphics[scale=0.5]{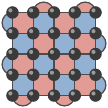}}}
 & $[[25,1,5]]$          & Ref.~\cite{Zen2025}        & all-to-all & 4 (4) & 4  & 0/0 & 16 & 10.93 & 8.53 & 54\\
 &                       & \textbf{ours}              & all-to-all & 4 (4) & 4  & 0/0 & \textbf{9} & \textbf{4.27} & 3.60 & \textbf{50}\\
\cline{3-11}
 &                       & Ref.~\cite{Colmenarez2026}, ME         & native & 24 (12) & 0  & 64/19936  & \textbf{7} & --   & -- & \textbf{40}\\
 &                       & Ref.~\cite{Colmenarez2026}, UEA        & native & 24 (12) & 0  & 64/24864  & 18         & --   & -- & 52\\
 &                       & \textbf{ours}, multi-size         & native & 24 (24) & 12 & 0/0        & \textbf{7} & 4.27 & 51.47 & 78\\
\cline{3-11}
 &                       & \textbf{ours}, multi-size         & hex    & 24 (20) & 10 & 0/0        & 10         & 4.27 & 32.53 & 64\\
\hline
\multirow{6}{*}{\raisebox{-0.4\height}{\includegraphics[scale=0.31]{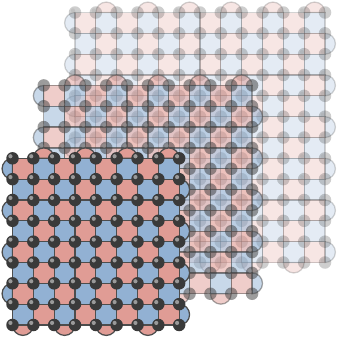}}}
 & $[[81,1,9]]$  & Ref.~\cite{Colmenarez2026}, ME   & native & 80 (40) & 0  & 256/345312/--/-- & \textbf{7} & --   & -- & \textbf{144}\\
 &               & Ref.~\cite{Colmenarez2026}, UEA  & native & 80 (40) & 0  & 256/422560/--/-- & 34         & --   & -- & 184\\
 &               & \textbf{ours}, multi-size   & native & 80 (80) & 40 & 0/0/0/0        & \textbf{7} & 17.07 & 196.00 & 284\\
\cline{3-11}
 &               & \textbf{ours}, multi-size   & hex    & 80 (72) & 50 & 0/0/0/0         & 10         & 17.07 & 155.73 & 268\\
\cline{2-11}
 & $[[121,1,11]]$ & \textbf{ours}, multi-size & hex & 120 (110) & 82  & 0/0/0/0/0   & 10 & 26.67 & 252.93 & 424\\
\cline{2-11}
 & $[[169,1,13]]$ & \textbf{ours}, multi-size & hex & 168 (156) & 109 & 0/0/0/0/0/0 & 10 & 47.73 & 379.60 & 640\\
\hline
\multirow{4}{*}{\raisebox{-0.4\height}{\includegraphics[scale=0.69]{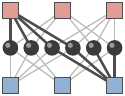}}}
 & $[[108,8,10]]$   & \textbf{ours}, $\ket{01{+}10^{4}}_L$ & all-to-all & 54 (54) & 405 & 0/0/0/0/0 & 66 & 809.80  & 0.00 & 2412\\
\cline{2-11}
 & $[[144,12,12]]$  & \textbf{ours}, $\ket{0^{12}}_L$ & all-to-all & 72 (72) & 78  & 0/0/0/0/0/0 & 14 & 164.27 & 0.00 & 868\\
 &                  & \textbf{ours}, $\ket{0101010^{6}}_L$ & all-to-all & 72 (72) & 72  & 0/0/0/0/0/0 & 15 & 105.60 & 0.00 & 828\\
\cline{2-11}
 & $[[186,10,14]]$  & \textbf{ours}, $\ket{0^{10}}_L$ & all-to-all & 93 (93) & 176 & 0/0/0/0/0/0/0 & 21 & 140.80 & 0.00 & 818\\
\hline
\end{tabular}
\caption{Logical-state preparation tasks. All cases prepare $\ket{0_L}$, except the $[[5,1,3]]$ 
code, which prepares $\ket{1_L}$, and the qLDPC codes, for which we specify each logical qubit's 
state ($0^{m}$ denotes $m$ consecutive zeros).  Gate set: H, S, X, Z, CNOT, CZ, ancilla measurement, 
and conditional X, Z. For the CSS codes, we use the $\ket{0_L}$ simplification of the fault 
tolerance criterion (Appendix~\ref{app:faulttol}), except for the $[[108,8,10]]$ code where the 
$\ket{+}$ qubit requires scoring both $X$- and $Z$-type residuals. ``anc.'' counts the ancillae 
available in the layout, and ``used'' those that the circuit actually uses. ``det.'' counts 
deterministic measurements and ``2q'' two-qubit gates. silent$_{j}$ lists the silent events of 
order $j$, comprising as many simultaneous faults. silent$_{1,2}$ are always enumerated. Higher 
orders are enumerated only when all lower orders vanish. Non-enumerated orders are marked by ``--''. 
Bold numbers mark the best entry for a given code and connectivity. ME (measurement encoding) and 
UEA (unitary encoding with ancilla) follow Ref.~\cite{Colmenarez2026}. The hex layout resembles 
the native layout, but drops each ancilla's coupling to the data qubit below it~\cite{McEwen2023}. 
``multi-size'' denotes a single program evolved simultaneously for all $d\in\{3,5,7,9\}$ for the 
colour code and surface code on the native layout, and $d\in\{3,5,7,9,11,13\}$ on the hex layout. 
For brevity, surface-code results at $d=7$ are omitted, as they follow the same trends as at 
other distances: On the native layout, our circuit, ME, and UEA have depths $7$, $7$, and $26$, 
respectively, with ME and UEA exhibiting silent faults. On the hex layout, our circuit has depth 
$10$. For the $[[7,1,3]]$ code, ``multi-size$^\dagger$'' denotes the circuit obtained from a 
separate multi-size search.  The circuits of Ref.~\cite{Colmenarez2026} have no detectors and 
therefore no heralded or wasted events.
}
\label{tab:states}
\end{table*}

Now that we have gone over the prompt and the verifier, we describe the agentic FunSearch. 
The FunSearch loop runs on several islands, each storing its own programs, partitioned into
clusters based on their scores. The search proceeds in generations, and in each generation, 
every island prompts the LLM with a few programs from its own archive, which includes its 
best program, one with a different number of measurements and a third drawn at random, with 
the last two present for diversity. Each program carries the verifier feedback, with its 
validity, fault metrics, and packed circuit (packed to maximize parallel qubit 
utilization). The feedback also marks which measurements serve as detectors and, for each 
silent event and ambiguous bin, gives the gates where the faults originate and the residuals 
they leave. The prompt also states the best score found so far across all islands as the frontier 
to beat. The LLM returns one new program, which is verified, scored, and added to that island. 
Additionally, the LLM returns scratchpad notes, which are shown to the LLM in the island's next 
prompt, allowing it to note down and persist positive ideas to reinforce, negative results to 
avoid, instructions for future generations, etc. The best programs periodically migrate between 
islands. To keep the discovery fully automated, every search starts cold with no seed programs. 
We arm the LLM with two tools: one verifies a circuit proposal and returns the same feedback the 
parents carry, and the other is a general Python tool with the circuit class, the targets and 
the verifier in scope, so that several proposals may be evaluated, ablated, and compared in one 
call. We use three LLM tiers, base, strong and elite. In most searches, the strong and elite 
tiers are GPT-6.1-sol with reasoning effort ``high'' and ``xhigh'', respectively. A cheaper 
model, GPT-6-luna with reasoning effort ``max'', serves as the base tier. Almost all proposals 
come from the strong tier, which also performs the cold start. For the Steane to Reed-Muller 
switch and the tier comparison of Fig.~\ref{fig:recordratewordcount}, the tiers are instead 
GPT-6-luna [high], GPT-6-luna [max] and GPT-6.1-sol [high].

\section{Results}
\subsection{Logical states}
\label{sec:states}

In this section, we use our agentic FunSearch for the preparation of several logical states
spanning several QEC codes, showcasing the capabilities of our approach. Tab.~\ref{tab:states} 
collects all results, on all-to-all connectivity and, for the surface code, also on its native 
and hex layouts.

\emph{Perfect code:--} The $[[5,1,3]]$ code~\cite{Laflamme1996} is non-CSS, 
which makes it a harder target than the CSS codes. Since its generators mix $X$ and $Z$ Paulis, 
none of them is automatically satisfied by the initial product state. Consequently, the 
simplification for CSS codes mentioned previously (detailed in Appendix~\ref{app:faulttol}), 
where the $Z$-type residuals can be ignored, no longer applies. We therefore score the full 
residual for every event. On all-to-all connectivity, we obtain a notably shallower circuit 
than the RL agent in Ref.~\cite{Zen2025}, using the same number of two-qubit gates.

\emph{Rotated surface code:--} The rotated surface code is a topological CSS code that, 
at distance $d$, encodes a single logical qubit in $d^2$ data qubits on a square 
lattice~\cite{BombinMartinDelgado2007,Fowler2012}. It has particular significance due 
to its favourable hardware layout, and high noise threshold~\cite{Fowler2012}, with 
several experimental demonstrations~\cite{Acharya2025,Bluvstein2024}. Here we consider 
it on all-to-all connectivity, the native layout where alternate sites on a nearest-neighbour
square lattice are data and ancilla qubits, and lastly the hex layout~\cite{McEwen2023}
which matches the native layout except that each ancilla's coupling to the data qubit 
below it is removed.

\begin{table*}[t]
\centering
\setlength{\tabcolsep}{3pt}
\small
\begin{tabular}{clllccccccc}
\hline\hline
 & code & gate & circuit & connectivity & anc. (used) & det. & silent$_{1/2}$ & \begin{tabular}{@{}c@{}}undecodable-\\[-3pt]cost\end{tabular} & depth & 2q  \\
\hline
\multirow{4}{*}{\raisebox{-0.4\height}{\includegraphics[scale=0.6]{Figures/code_perfect}}}
 & $[[5,1,3]]$  & $\overline{\mathrm{CNOT}}$ & Ref.~\cite{RyanAnderson2022}, CNOT3f & all-to-all & 2 (2) & 16 & 0 & -- & 84 & 65\\
 &              &                            & \textbf{ours}                   & all-to-all & 2 (2) & 4  & 0 & 0.00 & \textbf{16} & \textbf{29}\\
 &              &                            & \textbf{ours}, (A)              & all-to-all & 5 (5) & 5  & 0 & 0.00 & 9 & 31\\
 &              &                            & \textbf{ours}, (B)              & all-to-all & 5 (4) & 4  & 0 & 0.00 & 9 & 29\\
\hline
\multirow{9}{*}{\raisebox{-0.4\height}{\includegraphics[scale=0.51]{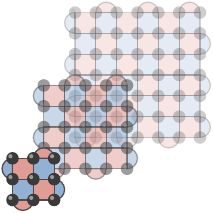}}}
 & $[[9,1,3]]$  & $\bar S$ & \textbf{ours}, single-size & \begin{tabular}{@{}c@{}}native\\[-3pt]($d=7$, centre)\end{tabular}
 & 48 (19) & 9 & 0 & 0.00 & 9 & 61\\[4pt]
 &              &          & \textbf{ours}, multi-size & native & 8 (5) & 12 & 0 & 0.00 & 41 & 62\\
\cline{4-11}
 &              &          & \textbf{ours}, multi-size & all-to-all & 16 (15) & 12 & 0 & 0.00 & 6 & 40\\
\cline{2-11}
 & $[[25,1,5]]$ & $\bar S$ & \textbf{ours}, single-size & \begin{tabular}{@{}c@{}}native\\[-3pt]($d=9$, centre)\end{tabular}
 & 80 (22) & 67 & 0/0  & 0.00 & 51 & 236\\[4pt]
 &              &          & \textbf{ours}, multi-size & native & 24 (14) & 100 & 0/0 & 0.02 & 191 & 344\\
\cline{4-11}
&              &          & \textbf{ours}, multi-size & all-to-all & 48 (33) & 33 & 0/0 & 0.00 & 9 & 128\\
\cline{2-11}
 & $[[49,1,7]]$ & $\bar S$ & \textbf{ours}, multi-size & all-to-all & 96 (65) & 65 & 0/0/0  & 0.00 & 10 & 277\\[4pt]
\hline
\multirow{3}{*}{\raisebox{-0.4\height}{\includegraphics[scale=0.6]{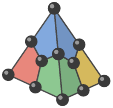}}}
 & $[[15,1,3]]$ & $\bar H$ & \textbf{ours} & all-to-all & 2 (2) & 3 & 0 & 0.00 & 37 & 46\\
 &              &          & \textbf{ours} & all-to-all & 6 (6) & 3 & 0 & 0.00 & 14 & 48\\
 &              &          & \textbf{ours} & all-to-all & 9 (9) & 3 & 0 & 0.00 & 10 & 48\\
\hline
\multirow{6}{*}{\raisebox{-0.4\height}{\includegraphics[scale=0.72]{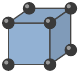}}}
 & $[[8,3,2]]$  & $\bar H_0$                 & Ref.~\cite{Kuehnke2025}, Fig.~3(a) & cube & 1 (1) & 1 & 0 & 32.33 & 19 & 13\\
 &              &                            & \textbf{ours}                 & cube & 1 (1) & 1 & 0 & \textbf{29.13} & \textbf{9} & \textbf{11}\\
\cline{3-11}
 &              & $\bar H_0\bar H_1$         & Ref.~\cite{Kuehnke2025}, Fig.~3(b) & cube & 1 (1) & 1 & 0 & 42.93 & 15 & \textbf{16}\\
 &              &                            & \textbf{ours}                 & cube & 1 (1) & 1 & 0 & \textbf{36.13} & \textbf{10} & \textbf{16}\\
\cline{3-11}
 &              & $\bar H_0\bar H_1\bar H_2$ & Ref.~\cite{Kuehnke2025}, Fig.~3(c) & cube & 1 (1) & 1 & 0 & 54.47 & 18 & \textbf{19}\\
 &              &                            & \textbf{ours}                 & cube & 1 (1) & 1 & 0 & \textbf{49.80} & \textbf{14} & 28\\
\hline
\end{tabular}
\caption{Logical gate design tasks. The columns are as in Tab.~\ref{tab:states}, with herald and 
waste replaced by the undecodable-cost of Eq.~\eqref{eq:score_gates}. The gate set is that of 
Tab.~\ref{tab:states}, extended by S$^\dagger$, Y, CY for the perfect-code 
$\overline{\mathrm{CNOT}}$, and by $\sqrt{X}$, CY for the $[[8,3,2]]$ code. For the surface-code 
$\bar S$ we consider three cases: all-to-all connectivity with two ancillae per check, 
its native layout (``native''), and at the centre of a larger one corresponding to the native 
patch of a higher-distance code, as specified. The embedded circuits come from a separate search 
per distance (``single-size''). The others come from a simultaneous search (``multi-size'') over 
$d=3,5$ on the native layout, and over $d=3,5,7$ on all-to-all connectivity. The multi-size program 
branches on distance instead of being size-general. The $[[8,3,2]]$ code uses the cube layout of 
Ref.~\cite{Kuehnke2025}, where the data qubits are coupled along the edges of a cube and a 
single ancilla at its centre is coupled to all of them. The subscripts of $\bar H$ label the 
logical qubits it acts on. All $[[8,3,2]]$ circuits show undecodable events as it is an error-detecting 
code. For the $[[15,1,3]]$ Reed-Muller and surface codes, the $X$ and $Z$ parts of event 
residuals are checked separately against Eq.~\eqref{eq:criterion} (Appendix~\ref{app:faulttol}). 
The undecodable-cost of the CNOT3f from Ref.~\cite{RyanAnderson2022} is marked ``--'' 
as we do not replicate the classical decoding and correction using the syndrome and flags as 
mentioned in their work. Our results for this logical CNOT on the $[[5,1,3]]$ code come from two 
separate searches, using two and five ancillae. For the latter, (A) and (B) are two results from 
the same search with the same score, but different two-qubit gate counts. Bold numbers mark the 
best entry for a given code, gate, and layout. }
\label{tab:gates}
\end{table*}

First, on all-to-all connectivity, we obtain a significantly shallower circuit than 
Ref.~\cite{Zen2025}. Besides this final circuit returned by our search, we also look into 
the program database and obtain an additional circuit (not shown) that sacrifices shallowness 
to achieve zero waste at depth $16$ and herald $4.27$.

Next, on the native layout, we run the search simultaneously (``multi-size'') on distances 
$d=3,5,7,9$, which is possible as the layout is scalable. Already in the first generation, 
the search obtains circuits that are valid and fault-tolerant with zero silent events at 
all distances. Five additional generations were spent on reducing the depth. The final program 
branches on distance, providing one circuit for $d=3$, and another for all other distances 
(Appendix~\ref{app:llmprogram_surface}). 
The former has depth $6$, while the latter has depth $7$ on all tested distances (up to $d=9$).
For $d\geq 5$, the circuit is similar to the standard measurement-based encoding (``ME'' 
in Ref.~\cite{Colmenarez2026} and Tab.~\ref{tab:states}); it measures the $X$ checks, 
and applies conditional $Z$ gates to correct the checks whose outcomes yielded $-1$.
At $d=3$, the circuit prepares the $X$ checks unitarily, similar to ``UEA'' in 
Ref.~\cite{Colmenarez2026}. However, at every distance, our circuit also measures the $Z$ 
checks in parallel, serving as fault detectors to catch the hook errors. For ME and UEA, 
our verifier shows silent events starting already at order one, i.e., from single faults, which 
is a result of our stricter fault tolerance criterion. They align the hook errors 
perpendicular to the logical $X$ operator $\bar X$, and thus, a decoder that can account for 
these hook errors in the subsequent QEC round could correct them. Our circuit, instead, uses 
its detectors to herald these events.

Lastly, we consider the hex layout~\cite{McEwen2023}. On running our search simultaneously 
for $d=3,5,7,9,11,13$, in the first generation we obtain fault-tolerant circuits with depths 
$7$, $11$, and $13$ at $d=3$, $d=5$ and $d\geq 7$, respectively. 
In the subsequent generations, the circuits at all distances shaved depth, and also lowered 
the herald and waste. In our circuit, since the ancilla of an $X$ check cannot reach 
all of its four data qubits directly as in the native layout above, it uses a neighbouring 
$Z$ check ancilla to reach all data qubits in its plaquette. The program branches on distance, 
providing a separate circuit for $d=3$.

\emph{Colour codes:--} Colour codes~\cite{BombinMartinDelgado2006,Bluvstein2024} 
are also topological CSS codes that, in two dimensions, support a transversal implementation of the 
entire Clifford group. We consider the two-dimensional triangular colour code family, spanning 
distances 3 through 9. The $d=3$ version, known as the Steane code, is the smallest
distance-three CSS code. It constitutes a simple test of our search, as fault-tolerant state
preparation on it has long been manually obtained. 

With all-to-all connectivity, we perform our search separately for each distance (``single-size''), 
and similar to the rotated surface code, also simultaneously for all distances (``multi-size''). 
The search progressed more slowly than that of the rotated surface code. In the multi-size search, 
fault-tolerant circuits were obtained already in the first generation for $d\leq 7$, but only in 
generation $11$ at $d=9$. The search did not find a general program for all distances. It took 
around $15$ generations for the depths to get to within $20\%$ of their final values. The last 
drop in depth came in generation $43$ for $d=5$. In the single-size runs, fault tolerance was 
achieved already in the first generation, and the depths converged to their final values sooner. 
At $d=5$ and $7$, our single-size search obtained notably shallower circuits, whereas at $d=9$, 
the multi-size search is shallower. For the $d=3$ Steane code, the multi-size search matches the 
depth $7$ of Ref.~\cite{Goto2016}, while another multi-size search (``multi-size$^\dagger$'') 
arrived at a shallower depth-$6$ circuit. 

\emph{qLDPC codes:--} High-rate quantum low-density parity-check (qLDPC) codes are the leading
family of QEC codes for low-overhead fault tolerance, as, unlike the codes above, they can encode
multiple logical qubits in a single block~\cite{Gottesman2014,Breuckmann2021,Bravyi2024}. We 
consider the $[[108,8,10]]$ and $[[144,12,12]]$ (gross) bivariate bicycle codes~\cite{Bravyi2024}, 
and the $[[186,10,14]]$ generalized toric code~\cite{Liang2025}. Despite the substantially longer 
verification times associated with their larger distances and gate counts, our approach 
successfully scales to these codes. It can independently prepare each logical qubit in its 
desired logical state. 

Beyond the $\ket{00\ldots 0}_L$ targets, we prepare $\ket{0101010^{6}}_L$ on the gross code and
$\ket{01{+}10^{4}}_L$ on the $[[108,8,10]]$ code. Note that the $\ket{+}$ qubit requires scoring 
both the $X$- and $Z$-type residuals. In every search, the first generation was already free of 
silent events up to order $d/2$, and subsequent generations reduced the depth. All circuits 
measure the $X$ checks with ancillae and apply $Z$ gates conditioned on the random outcomes.
A logical qubit in $\ket{1}$ is prepared by applying its logical $X$, and the $\ket{+}$ qubit 
by measuring its logical $X$ together with the $X$ checks. The program preparing $\ket{0101010^{6}}_L$ on the
gross code is presented in Appendix~\ref{app:llmprogram_gross010101000000}.

\subsection{Logical gates}
\label{sec:gates}

In this section, we run our agentic FunSearch to obtain physical circuits that implement 
logical gates on encoded qubits. The desirable scenario is that of a transversal gate, wherein 
the gate circuit acts separately on the constituent physical qubits of the code, thereby 
precluding the spread of faults. Unfortunately, no code admits a transversal implementation 
of a universal gate set~\cite{EastinKnill2009}, which calls for efficient and fault-tolerant 
circuit constructions for the non-transversal gates. We go over a few illustrative examples 
to show the strength of our approach. Tab.~\ref{tab:gates} collects all results.

\begin{table*}[t]
\centering
\setlength{\tabcolsep}{3pt}
\small
\begin{tabular}{cllcccccccc}
\hline\hline
 & code & task & connectivity & anc. (used) & det. & silent$_{1/2/3}$ & depth & herald & waste & 2q  \\
\hline
$\vcenter{\hbox{\includegraphics[scale=1.05]{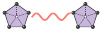}}}$\rule[-8pt]{0pt}{21pt}
 & $[[5,1,3]]$        & Bell pair            & all-to-all & 4 (4)   & 9  & 0 & 12 & 9.33 & 34.47 & 42\\
\hline\hline
 & code & task & connectivity & anc. (used) & det. & silent$_{1/2/3}$ & \multicolumn{2}{c}{\begin{tabular}{@{}c@{}}undecodable-\\[-3pt]cost\end{tabular}} & depth & 2q  \\
\hline
\multirow{2}{*}{\raisebox{-0.4\height}{\includegraphics[scale=1.05]{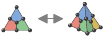}}}
 & $[[7,1,3]]$, $[[15,1,3]]$   & switch S $\to$ RM    & all-to-all & 2 (2)   & 4  & 0 & \multicolumn{2}{c}{0.00} & 20 & 40\\
 &                             & switch RM $\to$ S & all-to-all & 16 (8)  & 1  & 0 & \multicolumn{2}{c}{0.00} & 4  & 22\\
\hline
\multirow{2}{*}{\raisebox{-0.4\height}{\includegraphics[scale=1.05]{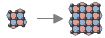}}}
 & $[[9,1,3]]\to[[25,1,5]]$    & grow $d=3\to5$     & native     & 24 (23) & 70  & 0/0  & \multicolumn{2}{c}{4.65} & 26 & 153\\
 & $[[25,1,5]]\to[[49,1,7]]$   & grow $d=5\to7$     & native     & 48 (48) & 181 & 0/0/0   & \multicolumn{2}{c}{0.12} & 32 & 465\\
\hline
\end{tabular}
\caption{Further applications: a logical Bell pair, code-switching, and code-growth. The gate 
set is as in Tab.~\ref{tab:states}. The Bell pair is a state preparation task, so its columns 
are as in Tab.~\ref{tab:states}. The code switches and growths are scored like logical gates by 
Eq.~\eqref{eq:score_gates}, so their columns mirror Tab.~\ref{tab:gates}. The Bell pair is the 
logical $(\ket{0_L0_L}+\ket{1_L1_L})/\sqrt2$ on two perfect-code encoded logical qubits. The 
Steane (S) $\to$ Reed-Muller (RM) switch switches a logical qubit in place from the S code to the 
RM code, and the RM $\to$ S switch switches and moves it from an RM block to a fresh S block. 
The growths take the rotated surface code from $d=3$ to $d=5$ and from $d=5$ to $d=7$. We 
use the fault tolerance criterion of the final code. For the growth, the $X$ and $Z$ parts 
of event residuals are checked separately against Eq.~\eqref{eq:criterion} (see 
Appendix~\ref{app:faulttol}). The growth results inevitably have non-zero undecodable-costs, 
as explained in Sec.~\ref{sec:applications}.
}
\label{tab:applications}
\end{table*}

\emph{CNOT on the perfect code:--} Unlike CSS codes, non-CSS codes do not permit 
a transversal CNOT, making its efficient and fault-tolerant implementation challenging. We use 
the $[[5,1,3]]$ perfect code, the smallest distance-3 non-CSS code, as a simple example. Since 
we have two logical qubit blocks, the fault tolerance criterion given by Eq.~\eqref{eq:criterion}
is applied separately to each encoded block. Reference~\cite{RyanAnderson2022} implements a logical 
CNOT between two perfect code blocks, representing two encoded logical qubits, with all-to-all 
connectivity. Their implementation comes in three variants, with all variants sharing a common 
backbone involving several single-qubit rotations and nine physical CNOTs between the blocks. 
The variant named ``CNOT3f'', containing
flagged checks, satisfies our fault tolerance criterion, and thus we use it for our benchmark. 
Reference~\cite{RyanAnderson2022} uses pieceable fault tolerance~\cite{Yoder2016}, where a
non-fault-tolerant gate is decomposed into small fault-tolerant pieces, interleaved with QEC cycles. 
Our circuits are fault-tolerant, and much shallower than CNOT3f (Tab.~\ref{tab:gates}). 
Our circuits measure only a total of four checks, two per logical block. In the two-ancilla 
circuit, the two ancillae flag each other's faults. In the five-ancilla circuit (A) each check 
has its own ancilla, two of these ancillae also flag each other, and the fifth ancilla 
simultaneously flags the remaining two ancillae. In (B), the fifth ancilla is unused, 
and the four check ancillae split into two pairs whose members flag each other, reducing the 
two-qubit gate count. Both searches achieve zero silent and undecodable events in the first 
generation, and achieve their final depths within three generations. 

\emph{Hadamards on the \texorpdfstring{$[[8,3,2]]$}{[[8,3,2]]} code:--} The $[[8,3,2]]$ 
code is a distance-$2$ error-\emph{detecting} 3D colour code. It encodes three logical qubits 
in eight data qubits located at the vertices of a cube. It supports a transversal non-Clifford 
gate, the CCZ~\cite{Bluvstein2024}, but it lacks a transversal 
Hadamard~\cite{Kuehnke2025}. Reference~\cite{Kuehnke2025} obtains logical Hadamard gates acting 
on up to three logical qubits simultaneously for this code, on its native cube connectivity 
with one flag qubit at its centre. We adopt this setting, and obtain shallower fault-tolerant 
circuits for these logical Hadamards, as shown in Tab.~\ref{tab:gates}. Since this is an 
error-detecting code, fault tolerance is enforced only by post-selection. In this case,
the undecodable-cost counts the ambiguous and heavy single faults.

\emph{Hadamard on the $[[15,1,3]]$ Reed-Muller code:--}
The $[[15,1,3]]$ quantum Reed-Muller code is a 3D topological colour code. As explained in 
Sec.~\ref{sec:applications}, it does not have a transversal logical Hadamard. 
Reference~\cite{Butt2024} implements it via code-switching to the Steane code, which 
admits a transversal Hadamard, and then switching back. Our circuits, with all-to-all 
connectivity, transversally apply $H$ to all data qubits and subsequently restore the 
Reed-Muller code by measuring $Z$ stabilizers and applying Pauli corrections conditioned on 
their outcomes. Three further measurements, whose outcomes are fixed by the earlier ones, 
act as detectors. We present the program and circuit for the two-ancilla case in 
Appendix~\ref{app:llmprogram_rmh}.

\emph{Phase gate on the rotated surface code:--} The phase gate lacks a 
transversal implementation on the rotated surface code. Previous implementations on the 
native layout temporarily use additional qubits beyond the logical qubit patch~\cite{Gidney2024Y,Hirai2026},
or require the injection of a resource state~\cite{Fowler2012}. We first attempted to 
find an ``in-place'' circuit with zero undecodable and forfeited events, using only the $d^2$
data qubits and $d^2-1$ ancillae of the logical patch. For this, we ran a simultaneous 
multi-size search for $d=3$ and $d=5$. Within the allocated budget of $60$ generations, 
it succeeded in obtaining a circuit with no undecodable events at $d=3$, but failed at 
$d=5$. Subsequently, the search succeeded for both distances when provided access to 
additional qubits by embedding the distance-$d$ logical patch at the centre of a larger 
distance-$(d+4)$ patch, both with the same native layout. In this case, we ran a separate 
search for each distance.

We also ran the search for all-to-all connectivity, simultaneously for $d=3,5,7$, with $d^2$ 
data qubits and $2$ ancillae per check. The search initially attempted a distance-general 
construction which was fault-tolerant only at $d=3$ and $5$. The first program fault-tolerant
at all three distances used one mechanism for $d=3$ and another for $d\geq 5$. The best program 
largely retained this structure, but tuned the detectors and gate order separately for 
$d=5$ and $d=7$.

\subsection{Further applications}
\label{sec:applications}

In this section, we tackle a curated list of tasks going beyond the state preparation and gate 
design objectives from the previous sections. Tab.~\ref{tab:applications} collects all results.

\begin{figure}[!htb]
\centering
\begin{subfigure}[t]{0.96\columnwidth}
\subcaption{}
\includegraphics[width=\columnwidth]{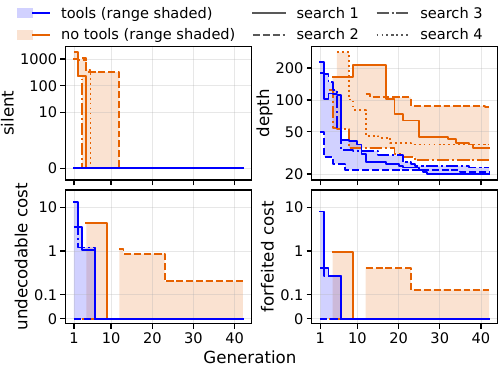}
\label{fig:storm_trajectory}
\end{subfigure}
\begin{subfigure}[t]{\columnwidth}
\subcaption{}
\includegraphics[width=0.999\columnwidth]{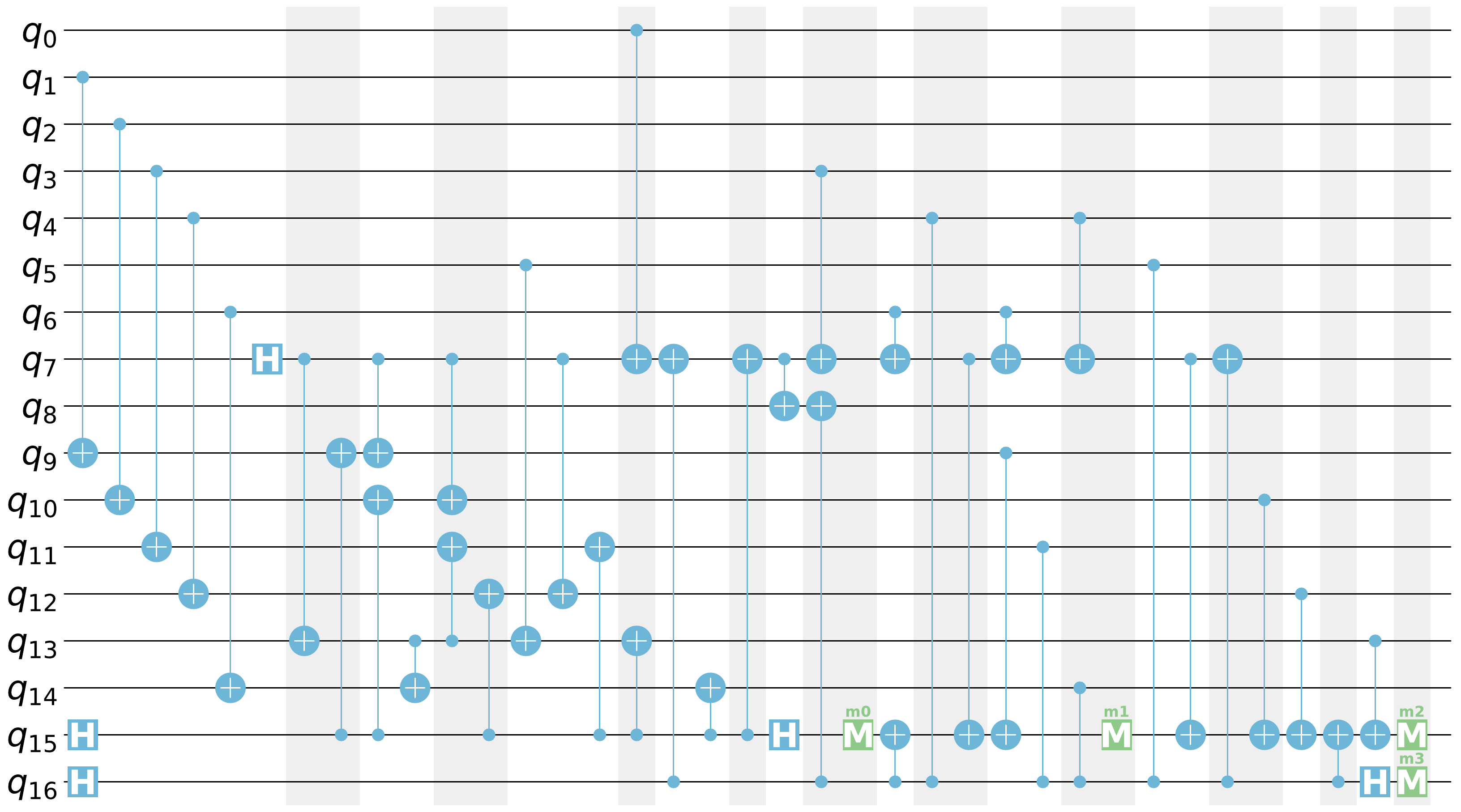}
\label{fig:storm_circuit}
\end{subfigure}
\caption{Steane to Reed-Muller code switch with all-to-all connectivity, and $2$ ancillae 
(Tab.~\ref{tab:applications}): (a) The fault tolerance metrics from Eq.~\eqref{eq:score_gates} of 
the best program up to each generation, for three agentic searches with tools (Sec.~\ref{sec:funsearch}, 
blue) and four non-agentic searches without tools (orange). Depth, undecodable- and forfeited-costs 
are shown only once silent drops to zero.  The range of scores for each kind of search is shaded. 
The agentic searches perform better on all counts (see Sec.~\ref{sec:disc}). Search 1 with tools 
is the one in Tab.~\ref{tab:applications}.  Since the ranking is lexicographic, a 
lower-priority metric can worsen while a higher-priority one improves, e.g., between generations 
$2$ and $3$, the undecodable-cost drops while the depth rises. (b) The best circuit, switching the 
Steane block on qubits $q_0$-$q_6$ to the Reed-Muller block on $q_0$-$q_{14}$, using 2 ancillae 
$q_{15}$-$q_{16}$. The green boxes denote the detectors. The subsequent QEC round uses their 
outcomes, along with the code-check syndrome, to correct the faults (see 
Sec.~\ref{sec:funsearch}). We shade alternate layers grey and white.}
\label{fig:storm}
\end{figure}

\emph{Logical Bell pair in the $[[5,1,3]]$ perfect code:--} Logical Bell pairs may be used 
as resource states for teleporting a logical qubit between two blocks, applying logical 
gates via gate teleportation~\cite{GottesmanChuang1999}, for Knill-style error correction~\cite{Knill2005,RyanAnderson2024},
etc. Here we prepare the logical Bell state 
$(\ket{0_L0_L}+\ket{1_L1_L})/\sqrt2$ in two logical qubits encoded in the perfect code.
As the perfect code is non-CSS, it lacks a transversal CNOT, which rules out the standard 
construction of a Bell state using a Hadamard followed by a CNOT. This is technically 
a state preparation task, with the target state being fixed by the code generators of 
both blocks, together with $\bar X\bar X=X^{\otimes 10}$ and $\bar Z\bar Z=Z^{\otimes 10}$, 
which are the usual generators of the Bell state elevated to the logical level. We enforce 
fault tolerance with the criterion in Eq.~\eqref{eq:criterion}, with $t=1$ per block. 

\emph{Code-switching between the $[[7,1,3]]$ Steane and the $[[15,1,3]]$ Reed-Muller code:--} 
Since no code has a transversal universal gate set, code-switching switches the encoding 
of a logical qubit between two codes with a complementary set of transversal gates. 
This provides access to a universal set of transversal gates, at the cost of the 
switching circuit. E.g., 3D colour codes admit transversal T and CNOT gates, while 2D 
colour codes admit transversal Hadamard and CNOT gates. Since the Hadamard, T, and CNOT 
together constitute a universal gate set, the logical Hadamard cannot be transversal in 
the former, while the T cannot be transversal in the latter. This is particularly relevant 
as a T magic state can be prepared in the Reed-Muller code, and then switched to the Steane 
code where the Cliffords are transversal. Reference~\cite{Butt2024} proposes fault-tolerant
code-switching between the $[[15,1,3]]$ quantum Reed-Muller code and the Steane code, which are 
3D and 2D colour codes, respectively. They perform the code-switching in-place in both 
directions, with the logical qubit staying in the same set of physical qubits throughout. 
References~\cite{Daguerre2025,Daguerre2025a} present a switch in a single direction, from the 
Reed-Muller code to the Steane code, where the logical qubit is teleported to a separate 
Steane block outside the initial Reed-Muller block. We consider both settings, switching 
from Steane to Reed-Muller in-place, and from Reed-Muller to Steane by transporting the 
logical qubit to a fresh Steane block. In both cases, the target is the Choi state of 
the logical identity from the initial code to the final code, defined suitably on the 
physical qubits. For the fault tolerance criterion, we choose $t$ to coincide with that 
of the final code. This happens to be $t=1$ in our case, but in general, the final code 
may have a different distance than the initial code. The results are shown in 
Tab.~\ref{tab:applications}. Both circuits have zero silent, undecodable and forfeit 
costs. The Steane to Reed-Muller circuit (Fig.~\ref{fig:storm_circuit}) uses its ancilla 
measurements only as detectors. Fig.~\ref{fig:scratchpad} shows one representative 
scratchpad note from each of the three LLM tiers (GPT-6-luna [high], GPT-6-luna [max], 
and GPT-6.1-sol [high]) during the search. We also highlight the benefits of an agentic 
FunSearch with tools in Fig.~\ref{fig:storm}(a), which shows that searches with tools 
consistently achieve better circuits within the same generational budget. The Reed-Muller 
to Steane circuit reads the qubits outside the Steane face in the $X$ basis, moves the 
Steane face to the fresh block with two CNOTs per qubit, and restores the signs of the 
Steane $X$ checks with conditional $Z$ gates. We present the program and circuit for the 
Reed-Muller to Steane switch in Appendix~\ref{app:llmprogram_rmtos}. 

\emph{Rotated surface code-growth:--} Growing the code distance of an encoded logical 
qubit forms a key component of magic state cultivation~\cite{Gidney2024,GoogleQAI2025Cultivation,
Sahay2026} and modular quantum computing~\cite{Marqversen2026}. In both cases, a resource 
state such as a magic state or a Bell pair is injected into a small-distance code and then 
the code is grown to the desired distance. In our implementation, as in the code-switching 
task above, the target is the Choi state of the logical identity. Fault tolerance is verified 
by using the criterion for the final code. We note that code-growths are inherently 
fault-tolerant only under post-selection. Unlike a logical gate or a code switch at a fixed 
distance, in a growth from distance $d_i$ to $d_f$, an undetected logical error at the 
input can have a weight $d_i<d_f$, which is lower than the distance $d_f$ used for the 
fault tolerance criterion. With $t_f=\lfloor (d_f-1)/2\rfloor$, zero silent events then 
require $d_i>t_f$, to avoid getting enumerated. Even then, assuming odd distances, since 
$d_i<2t_f$, the weight-$d_i$ input logical error can split into two events of at most $t_f$ 
simultaneous faults each, both of which are enumerated. As their combination is undetected 
with trivial syndrome and detector pattern, and the syndrome and pattern of an event are 
the XOR of those of its constituent faults (see Appendix~\ref{app:faulttol}), they have the same 
detector pattern and syndrome, yet their residuals differ by a logical operator. Their bin 
is thus ambiguous, and both are discarded. The smaller event has as few as $d_i-t_f$ faults, 
when the other has $t_f$. Thus, a growth is fault-tolerant only under post-selection, with 
the discards starting at events of order $d_i-t_f$ or lower. Nevertheless, we use the event 
classification and scoring of the logical gates, rather than those of state preparation. 
This is because the latter discards indiscriminately on encountering the pattern of any 
heavy event residual. Instead, the former discards a shot only if its bin is ambiguous, 
i.e., only if it cannot be corrected, reducing the wastage of shots.

We consider the growths $d=3\to5$ and $d=5\to7$. Both circuits in Tab.~\ref{tab:applications} 
have zero silent events, and their first discards occur at order $d_i-t_f$ mentioned above, 
$1$ and $2$, respectively. The $d=3\to 5$ search was not fully converged within its 
allocated budget, and it was still improving in its final ($34$th) generation. For the 
$d=5\to 7$ case, the search was almost stalled by the $33$rd generation. Both 
circuits repeatedly measure the checks of the larger code, apply Pauli gates conditioned 
on the outcomes, and use the deterministic outcomes as detectors.

\section{Discussion}
\label{sec:disc}

Here we highlight two key features of our agentic FunSearch, namely, the choice of 
LLMs, and the role of tool-use.

\begin{figure}[!htb]
\centering
\includegraphics[width=\columnwidth]{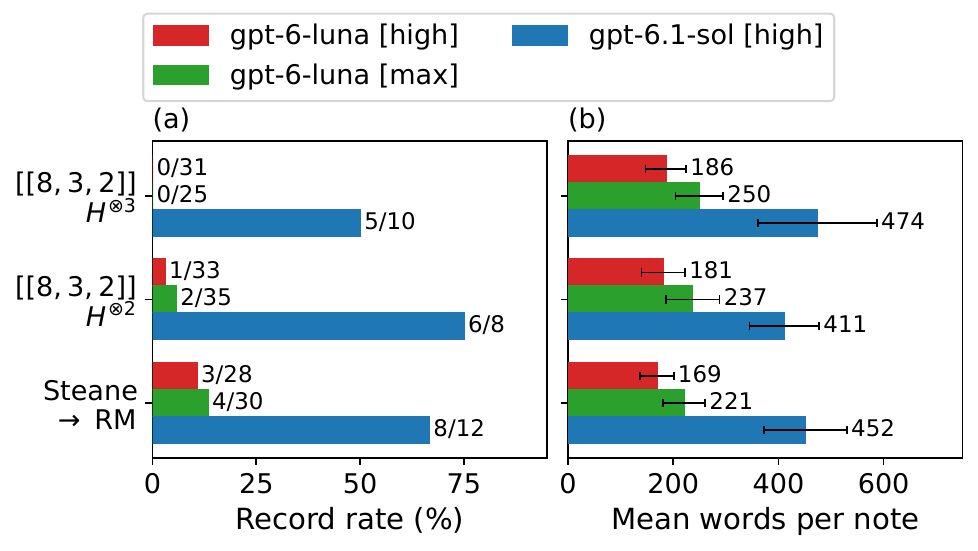}
\caption{(a) The record rate of each LLM tier, i.e., the percentage of its programs 
that improve the best score found so far by any tier. Both panels consider all programs 
up to the generation of the best program, excluding the cold-start generations, which do not 
call the base tier. Next to each bar, we show the number of records / generated programs. 
We show three targets, each from a single search: the logical Hadamard on logical qubits 
$0,1,2$, and $0,1$ on the $[[8,3,2]]$ code, and the Steane to Reed-Muller code switch. 
Except for the Steane to Reed-Muller switch, the rest were obtained separately from 
those in Tab.~\ref{tab:gates} by allocating more proposals to GPT-6-luna. The elite 
tier achieves a significantly higher record rate. (b) Mean word count (error bars 
represent $\pm 1$ standard deviation) per scratchpad note for all programs (not just 
record candidates). For the cases shown here, increasing GPT-6-luna's reasoning effort 
from high to max raises word count by $\approx 30\%$. Switching to a better LLM, 
GPT-6.1-sol, at the same high reasoning effort, more than doubles the word count. 
The higher word count suggests more hypotheses tested and ideas left behind. We present 
the notes for the Steane to Reed-Muller switch in Fig.~\ref{fig:scratchpad}.
}
\label{fig:recordratewordcount}
\end{figure}

We recall that the standard FunSearch approach relied on relatively weak LLM calls over many 
generations. This hinged upon the premise that several mutations of an evolving artifact 
will eventually introduce high-performing features, even when the individual proposals are 
of limited quality. However, we find that a few generations driven by frontier LLMs suffice 
to rapidly and reliably find performant circuits within a limited budget. We believe that 
this follows from the improved proposal quality of modern frontier LLMs and their proficiency 
in agentic tool use, allowing the search to escape local minima in the fitness landscape. This 
is illustrated in Fig.~\ref{fig:recordratewordcount}(a), where we find that the elite LLM, 
which is the strongest model we use (GPT-6.1-sol), is far more likely to produce new 
records in the scored metrics. Associated with these observations, as shown in 
Fig.~\ref{fig:recordratewordcount}(b), we also note that the elite LLM consistently 
leaves longer and more elaborate scratchpad notes, despite being provided the same prompt 
as the remaining tiers. 

Moreover, the agentic nature of our FunSearch substantially enhances its performance. With
tool use, the LLM extensively tests and refines its proposals based on the verifier's output, 
instead of having to reason about fault propagation without any evidence. To quantify this, 
we ran the Steane to Reed-Muller switch search of Fig.~\ref{fig:storm_trajectory} three times 
with tools and four times without, using the same LLM tiers, prompt (except for tool 
descriptions), and generational budget. All three searches with tools have no silent events 
from the first generation. Without tools, reaching zero silent events takes several generations, 
and the circuits are consistently deeper. The LLM itself mentions the lack of tools in its 
scratchpad notes. Roughly half of the notes written without tools state that the submitted 
circuit is untested, e.g., ``a reasoned equivalence, not a measured result'', and the 
proposals are defensive and limited to smaller changes that can be argued as fault-tolerant 
without an explicit verification, such as relabeling qubits by a code automorphism or 
swapping commuting gates. A representative note is shown in Fig.~\ref{fig:scratchpad_notools} 
in Appendix~\ref{app:scratchpadnotes}. Lastly, we remark that the run time remains largely 
unchanged irrespective of tool use, presumably because in the absence of any verification 
the LLM spends more time textually reasoning about the effectiveness of each change. 

\section{Conclusion and Outlook}
\label{sec:conc}

We have presented an agentic evolutionary search, in which an LLM writes Python programs 
corresponding to dynamic Clifford circuits for various quantum computing tasks. The key 
result is that modern frontier LLMs, armed with tool use and appropriate feedback 
reflecting our objectives, validity and fault tolerance metrics of the circuit, are 
proficient in autonomously obtaining performant circuits. To this end, we have considered 
a curated set of examples spanning logical state preparation and gate design, as well as 
logical routines such as code-switching and code-growth. All discovered circuits satisfy 
our fault tolerance criterion, matching or improving upon existing benchmarks (based on 
our score) wherever a direct comparison is possible. In fact, our search can be applied to 
any target that can be specified as a Clifford map on the qubits by using the Choi state. 

Our approach provides a general template that can be extended in several directions. 
While we have focused on a particular hierarchy of fault metrics, minor modifications to 
the verifier and prompt would allow us to target any desired objective, e.g., different 
noise models and fault tolerance criteria, measurement-free unitary circuits, explicit 
optimization of two-qubit gate counts, etc. The size of the circuits that our approach can 
currently handle is determined by the speed and memory requirements of the verifier, since 
the number of fault combinations it must check grows rapidly with the code distance and 
circuit size. We already reach distance $14$ for logical state preparation, succeeding in 
several qLDPC codes, and distance $7$ for logical gates. Improvements in the verifier 
would push this limit further. Another promising direction is an extension to non-Clifford 
gates, which are needed, e.g., for magic state preparation. This entails a redesign of 
the verification modules as non-Clifford gates generally turn Pauli faults into non-Pauli 
residuals, but it appears feasible for sufficiently small circuits. Our approach is not 
tied to the layouts considered here, and it accommodates arbitrary connectivities 
specified in the prompt. In fact, one may add another layer of LLM-guided evolutionary 
search, which optimizes the compilation strategy under hardware-specific constraints 
on connectivity, resources, or operations.

More broadly, we demonstrate the potential of LLM-based agents as a general paradigm for 
designing quantum circuits for various tasks, showcasing the strengths of agentic program 
synthesis coupled with rigorous verification. 

\emph{Note:--} During the final stage of writing this manuscript, three related works on
LLM-based search for quantum circuits appeared on the arXiv~\cite{Gu2026,Khattar2026,Crichigno2026}. 
References~\cite{Gu2026,Khattar2026} optimize Toffoli and qubit counts of circuits 
computing certain functions on logical qubits, leaving fault tolerance to future extensions, 
and Ref.~\cite{Crichigno2026} targets certain compilation problems. We instead search for 
Clifford circuits at the level of physical qubits, with exhaustively verified fault tolerance.

\let\oldaddcontentsline\addcontentsline
\renewcommand{\addcontentsline}[3]{}

\begin{acknowledgments}
This research is part of the Munich Quantum Valley, which is supported by 
the Bavarian state government with funds from the Hightech Agenda Bayern Plus.
\end{acknowledgments}

\section*{AI Use Disclosure}
We used Claude Code (Claude Opus 5.5, 5.1, 5, and 4.8) to 
help write the code, and it was instrumental in speeding up the exhaustive fault 
enumeration. It also suggested the technique of Ref.~\cite{Kliuchnikov2023}. All 
methods and interpretation are the authors' own, or cited. We tested the verification 
and fault-tolerance code on small circuits that can be checked by hand. We reviewed 
all AI-generated content and take full responsibility for it.

\appendix

\section{Logical gate target: logical Choi state}
\label{app:gatechoi}

We briefly review the Choi state construction for the representation of logical gates. 
A logical gate $\bar U$ must map every input codeword to its image under the logical gate 
operation. However, the channel-state duality~\cite{Jamiolkowski1972,Choi1975}, 
which represents a gate by its Choi state~\cite{NielsenChuang1997,GottesmanChuang1999,Peham2023}, 
converts this into a single target state in an enlarged Hilbert space. Hence, the logical 
gate design task inherits the technical machinery already used for state preparation. To 
illustrate the idea, we first consider a single physical qubit. We pair it with a reference 
qubit in the Bell state $\ket{\Phi}=(\ket{00}+\ket{11})/\sqrt{2}$, where the first qubit 
is the physical qubit and the second one is the reference qubit. We apply a map $U$ to the 
physical qubit, giving $(U\otimes I)\ket{\Phi}=\sum_j(U_{j0}\ket{j0}+U_{j1}\ket{j1})/\sqrt{2}$. 
The reference qubit is purely a bookkeeping device, disconnected from the physical qubits, 
and absent in the actual physical device where the unitary is to be implemented. Since it is 
ntangled with the initial physical qubit, it provides a memory of the initial state. As such, 
any map that produces this Choi state agrees with $U$. In the stabilizer formalism, the Bell state 
$\ket{\Phi}$ is fixed by $X\otimes X$ and $Z\otimes Z$, so $(U\otimes I)\ket{\Phi}$ is fixed 
by $UXU^\dagger\otimes X$ and $UZU^\dagger\otimes Z$. For a Clifford $U$, each image is again 
a Pauli operator up to a sign, $UXU^\dagger=\pm P_X$ and $UZU^\dagger=\pm P_Z$. The unsigned 
operators $P_X\otimes X$ and $P_Z\otimes Z$ become the new rows of the symplectic part $(x|z)$, 
and the two signs enter the corresponding sign bits. Moving now to several \emph{logical} 
qubits, we add $k$ reference qubits, each forming a Bell pair with a logical qubit of the 
code, and the same argument applies, with logical unitaries $\bar U$ in place of the original 
physical unitaries. In the stabilizer formalism, this amounts to elevating the stabilizers 
representing the Bell state, $XX$ and $ZZ$, to the logical level for the code qubits, 
$\bar X_j\otimes X_{r_j}$ and $\bar Z_j\otimes Z_{r_j}$, where $j=1,\dots,k$ denote the 
$k$ logical qubits and $r_j$ denote the corresponding entangled reference qubits. In 
addition, we append the $n-k$ code stabilizers. The target state is the state reached 
when $\bar U$ acts on the code block alone. It is stabilized by the code stabilizers 
together with $\bar U\bar X_j\bar U^\dagger\otimes X_{r_j}$ and 
$\bar U\bar Z_j\bar U^\dagger\otimes Z_{r_j}$. Because the reference is entangled with every 
logical state at once, reaching this single stabilizer state certifies the gate on all 
possible codewords.

\section{Fault tolerance}
\label{app:faulttol}

Here we describe our fault tolerance criterion and its implementation in detail.

As mentioned in Sec.~\ref{sec:funsearch}, each fault $f$ is propagated from its birth 
location to the end of the circuit to obtain its residual $E_f$, which is a Pauli on 
the data qubits. While the faults are tracked and propagated on all qubits, including 
the ancillae, at the end the residual is read only on the data qubits. The residual of 
higher-order events, comprising $\geq 2$ simultaneous faults, is obtained from that of 
its constituent single faults as described later in this section. Each event is judged 
by its residual using the reduced weight 
$\mathrm{wt}_{\mathcal{G}}(E_f)=\min_{G\in \mathcal{G}} \mathrm{wt}(E_fG)$. This involves 
two objects which we first introduce here, and obtain in more detail below: (a) The group 
$\mathcal{G}$ of Paulis on the data qubits that stabilizes the target. For any 
$G\in\mathcal{G}$, the residuals $E_f$ and $E_fG$ act identically on the target, even 
though their weights, $\mathrm{wt}(E_f)$ and $\mathrm{wt}(E_fG)$, may differ. The reduced 
weight provides an unambiguous answer. (b) A set of Paulis $\mathcal{R}$, against which 
we compute the syndrome of the residual, for efficiently checking the reduced weight 
bound in Eq.~\eqref{eq:criterion} without enumerating all elements of $\mathcal{G}$. 
We recall that the syndrome of a Pauli $P$ against another set of Paulis 
$\mathcal{P}=\{\mathcal{P}_1,\ldots, \mathcal{P}_{N_{\mathcal{P}}}\}$ is a binary vector 
$\text{syn}_{\mathcal{P}}(P)=[\langle P,\mathcal{P}_1\rangle ,\ldots, \langle P,\mathcal{P}_{N_{\mathcal{P}}}\rangle ]$ 
such that $P\mathcal{P}_l=(-1)^{\langle P,\mathcal{P}_l\rangle }\mathcal{P}_lP$. That is, 
$\langle P,\mathcal{P}_l\rangle =0$ if $P$ commutes with $\mathcal{P}_l$, and $1$ if it anticommutes. 
It satisfies two properties: (1) $\text{syn}_{\mathcal{P}}(P)=\text{syn}_{\mathcal{P}}(PQ)$ 
when $\text{syn}_{\mathcal{P}}(Q)=0$, i.e., $[Q,\mathcal{P}_l]=0$ for all $l$. (2) If 
$\text{syn}_{\mathcal{P}}(P)=\text{syn}_{\mathcal{P}}(P')$, then $\text{syn}_{\mathcal{P}}(PP')=0$ 
and hence, $P=P'Q$, where $\text{syn}_{\mathcal{P}}(Q)=0$. As such, $\text{syn}_{\mathcal{P}}(P)$ 
identifies $P$ modulo the Paulis that commute with every element of $\mathcal{P}$. 
The usual syndrome is calculated against the $n-k$ code checks. We use the syndrome of the 
residual against $\mathcal{R}$, calling it the extended syndrome, and choose $\mathcal{R}$ 
such that $\text{syn}_{\mathcal{R}}(Q)=0$ iff $Q\in\mathcal{G}$ up to phases (henceforth, 
a ``syndrome'' refers to the usual syndrome against the code checks). Hence, 
following Eq.~\eqref{eq:criterion}, $\mathrm{wt}_{\mathcal{G}}(E_f)\leq i$ iff 
$\text{syn}_{\mathcal{R}}(E_f)$ matches that of some Pauli of weight at most $i$. 
Therefore, it is sufficient to list the extended syndromes of all Paulis of weight at 
most $i$, and check if $\text{syn}_{\mathcal{R}}(E_f)$ belongs to that list. Below we 
describe the choice of $\mathcal{G}$ and $\mathcal{R}$ for the various tasks considered 
in this work.

First, we consider logical state preparation. In this case $\mathcal{G}=\mathcal{S}$, 
the stabilizer group of the target state, since any element of $\mathcal{S}$ acts trivially 
on the target state. $\mathcal{S}$ contains the $n-k$ code checks as well as the $k$ 
logical operators that pin the target logical state. The extended syndrome is computed 
against $\mathcal{R}$ containing these $n$ generators, as $\mathrm{syn}_{\mathcal{R}}(Q)=0$ 
iff $Q\in\mathcal{S}$. Hence, the extended syndrome uniquely identifies the residual 
$E_f$ up to an element of $\mathcal{S}$~\cite{Zheng2018,GottesmanBook2026}. The QEC round 
measures the code checks, and using standard minimum-weight decoding, it applies the 
lightest correction $C$ that has the same syndrome as $E_f$. Since every $S\in\mathcal{S}$ 
commutes with the code checks, $E_fS$ has the same syndrome as $E_f$, and thus 
$\mathrm{wt}(C)\leq\min_{S\in\mathcal{S}}\mathrm{wt}(E_fS)=\wtS(E_f)$. Furthermore, since 
$C$ has the same syndrome as $E_f$, $CE_f$ has zero syndrome, so it must either be the 
product of some code checks, or a logical operator. As per the bound in Eq.~\eqref{eq:criterion}, 
in the case where $\wtS(E_f)\leq i$, we can conclude that $\wtS(CE_f)\leq 2\wtS(E_f)\leq 2i\leq 2t<d$. 
While any logical operator $L$ has $\mathrm{wt}(L)\geq d$, the reduced weight of those 
within $\mathcal{S}$ is zero as $\wtS(L)=\min_{S\in \mathcal{S}} \mathrm{wt}(LS)=\mathrm{wt}(LL)=0$. 
However, for a logical operator $L'$ outside $\mathcal{S}$, i.e., a logical error, we have 
$\wtS(L')=\min_{S\in \mathcal{S}} \mathrm{wt}(L'S)=\mathrm{wt}(L'')$ where $L''$ 
is again a logical operator outside $\mathcal{S}$. Hence, $\wtS(L')\geq d$. Therefore, 
$CE_f$ cannot be a logical error and must belong to $\mathcal{S}$, i.e., it is a product of code 
checks, possibly times logical operators that pin the state, and acts trivially on the 
target state. 

Now, we consider logical gate design. For a logical gate $\bar U$, the target is its Choi 
state. A residual Pauli on the data qubits should preserve the code space and also act 
trivially on the logical space. Hence, we use $\mathcal G=\mathcal S_\text{code}$, the 
group generated by the $n-k$ code checks (henceforth denoted as $\{g_m\}$ where 
$m=1,\ldots,n-k$). While the Bell generators stabilize the Choi state, they are not included 
as they act on the reference qubits, and they contain logical operators which, when 
multiplied with a residual on the data qubits, alter the action of the gate. Now, as in 
the case of logical state preparation, a natural starting point for $\mathcal{R}$ is the code 
checks $\{g_m\}$, as all of $\mathcal{G}$ has zero syndrome against them. But this is 
not sufficient as there are other Paulis which have zero syndrome against the code checks 
as well. Specifically, a residual has zero syndrome against $\{g_m\}$ when it commutes 
with every $g_m$, i.e., it lies in the normalizer $N(\mathcal S_\text{code})$. 
This contains not only $\mathcal S_\text{code}$ itself, but also all the logical Pauli 
operators, $k$ each of $\bar{X}$ and $\bar{Z}$. This is an issue as a syndrome against 
$\mathcal{G}$ alone would let logical errors pass through. E.g., for a Pauli $P$ with 
$\mathrm{wt}(P)\leq t$, the residuals $E_f=P\bar X_j$ and $E_f=P$ have the same syndrome 
against $\{g_m\}$, and our verifier would then judge them similarly. However, on 
encountering $E_f=P\bar X_j$, the QEC round would apply the lightest correction $C$ with 
this syndrome, so $\mathrm{wt}(C)\leq\mathrm{wt}(P)\leq t$. Then $CP$ has zero syndrome 
with weight at most $2t<d$, so it must lie in $\mathcal{S}_{\text{code}}$. Hence, we are 
left with $CE_f=(CP)\bar X_j$, which is a logical error that would alter the gate action. 
Therefore, we calculate the extended syndrome against the code checks $\{g_m\}$, along 
with the $2k$ logical operators
\begin{equation}
  \mathcal R \;=\; \{g_m\}_{m=1}^{n-k} \ \cup\  \{\bar U\bar X_j\bar U^\dagger\}_{j=1}^{k} \ \cup\ 
  \{\bar U\bar Z_j\bar U^\dagger\}_{j=1}^{k},
  \label{eq:R}
\end{equation}
where the last two terms are the $2k$ logical operators of the output code, mapped by the
target gate. This extended syndrome is a length-$(n+k)$ binary vector. As $\bar U$ is a 
logical Clifford, the mapped rows are still logical Pauli operators. We use these mapped 
rows instead of $\bar X_j$ and $\bar Z_j$ because they already constitute a part of 
the logical Bell generators of the target Choi state (Appendix~\ref{app:gatechoi}). 
$\mathcal R$, as constructed here, is such that a residual with zero extended syndrome 
against it must be a code check, and not a logical operator. This is because commuting 
with the $\{g_m\}$ means that the residual is not detected by the following QEC round, so 
it is a product of any number of code checks and logical operators. Commuting with all 
$2k$ logical operators means that it acts as an identity on the logical qubits. Hence, it 
can only be a code check, and the syndrome against $\mathcal R$ fixes a residual up 
to a code check. In the example above, the two residuals differing by $\bar X_j$ are now 
distinguished by their commutator with the last $2k$ rows of $\mathcal R$ containing the 
logical operators. 

Two cases need a small adaptation. For a code switch, the residual is judged on the output 
code's qubits, and $\mathcal R$ is built from the output code. For a gate on several code 
blocks, such as a logical CNOT between two blocks, each block is followed by its own QEC 
round, so Eq.~\eqref{eq:criterion} is applied to each block separately.

For an error-\emph{detecting} code with distance two, such as the $[[8,3,2]]$ code used in 
Sec.~\ref{sec:gates}, we have $t=0$. Hence, the fault tolerance criterion concerns only 
detection, as nothing can be corrected. A single fault is then fine only if its residual is a
stabilizer. If the residual has a non-zero syndrome, the following QEC round detects it, and 
the fault is undecodable. If the syndrome is zero, the residual may be a stabilizer, in which 
case a non-trivial pattern makes it forfeited, or the residual may be a logical error, in 
which case it is silent when the pattern is trivial, and undecodable when the pattern is 
non-trivial.

We remark that CSS codes admit some simplifications. We start with the preparation of $\ket{0_L}$ 
or $\ket{+_L}$. To this end, write each event residual as $E=E_XE_Z$, ignoring phases. Since 
every code check is either purely $X$- or purely $Z$-type, the syndrome of $E_Z$ ($E_X$) comes 
only from the $X$-type ($Z$-type) checks. We therefore assume a CSS decoder in the subsequent 
QEC round, which corrects the two parts independently. Its $X$- and $Z$-type corrections, $C_X$ 
and $C_Z$, have the same syndrome as $E_X$ and $E_Z$, respectively. So, $C_ZE_Z$ is a $Z$-type 
Pauli that commutes with every check. Hence, it is a product of $Z$-type checks, possibly times 
logical $Z$ operators, all of which leave $\ket{0_L}$ unchanged. Therefore, $Z$-type residuals 
can be ignored when preparing $\ket{0_L}$~\cite{Goto2016,Zen2025,Colmenarez2026}. A similar 
argument applies to the preparation of $\ket{+_L}$, where we can ignore the $X$-type event 
residuals. 

A similar simplification applies to logical gates and other applications on CSS codes. We 
use it for the Hadamard on the Reed-Muller code, the phase gate on the surface code, and 
code-growth. We again assume a CSS decoder in the subsequent QEC round, which corrects 
$E_X$ and $E_Z$ independently. Since the logical $\bar{Z}$ ($\bar{X}$) operator is purely 
$Z$-type ($X$-type), and of weight at least $d$, $E_X$ and $E_Z$ can each be corrected 
independently up to weight $t$. We therefore judge the two parts independently. An event of 
order $i$ is light when $E_X$ has the same syndrome against $\mathcal R$ (Eq.~\eqref{eq:R}) 
as some $X$-type Pauli of weight at most $i$, and $E_Z$ as some $Z$-type Pauli of weight at 
most $i$. For example, a single fault leaving $X_aZ_b$ on two data qubits is generally heavy 
under Eq.~\eqref{eq:criterion}, but each of its parts has weight one and is corrected. 
For both parts, the syndrome is taken against all of $\mathcal R$. Note that, unlike for 
state preparation, the $2k$ logical Bell stabilizers in $\mathcal R$ need not be pure, 
e.g., $\bar U\bar X\bar U^\dagger=\bar Y$ for the phase gate. Such stabilizers respond to 
both $E_X$ and $E_Z$, so it is not obvious that the two can be judged independently. To 
see how this is not an issue, note that the correction succeeds when $(E_XC_X)(E_ZC_Z)$ is 
a stabilizer, i.e., has zero syndrome against $\mathcal R$. Since every stabilizer of a CSS 
code is a product $S_XS_Z$ of a pure $X$-type and a pure $Z$-type one, this holds iff $E_XC_X$ 
and $E_ZC_Z$ are each stabilizers.

Now, we briefly go over our argument for how the idling faults are indirectly captured by 
our approach. Consider a qubit waiting between two consecutive operations. Nothing acts on 
that qubit in between, so a Pauli inserted anywhere inside the window is conjugated by 
precisely the same gates as one inserted immediately after the earlier operation. Whenever 
either end of the idling window is a gate, the idling fault therefore reproduces one of the 
fault Paulis already enumerated there, and our earlier discussion applies. As a result, 
idling faults cannot invalidate the fault tolerance. Since the idle time of the qubits, 
and consequently the possibility of idling faults, is commensurate with the depth of the 
circuit, we explicitly minimize the depth. Reducing depth is also beneficial for reducing 
the runtime of the circuit.

The fault tolerance criterion given by Eq.~\eqref{eq:criterion} considers events of every 
order up to $t$, i.e., all sets of up to $t$ simultaneous faults, the number of which increases 
combinatorially with increasing order and circuit size. We efficiently enumerate all the events 
and their patterns as follows: (a) A fault is a Pauli inserted immediately after a non-Pauli 
gate or at the input of a qubit, or a flipped measurement outcome. The Pauli gates are tracked 
in software and carry no faults~\cite{Knill2005}. Ignoring phases, which do not affect a residual's 
weight or syndrome, a Pauli $P$ is specified by two bit vectors $x$ and $z$ marking the qubits 
where it has an $X$ or a $Z$ component. The product of two Paulis $P_1$ and $P_2$ has the bit 
vectors $x_1\oplus x_2$ and $z_1\oplus z_2$. All faults are propagated to the end of 
the circuit, and we accumulate the two things required to quantify fault tolerance: 
the residual that a fault leaves on the data qubits, and its pattern (set of flipped 
detectors). Faults are propagated by using the fact that a fault Pauli $F$ preceding a 
Clifford gate $U$ is equivalent to the Pauli $UFU^\dagger$ following the gate. This operation 
acts on $F$'s bits $\GF$-linearly by XOR and swap, e.g., $x_b\leftarrow x_b\oplus x_a$ and 
$z_a\leftarrow z_a\oplus z_b$ for a CNOT from qubit $a$ to qubit $b$, $x_q\leftrightarrow z_q$ 
for $H$ on qubit $q$, etc. A $Z$ measurement of qubit $q$ flips its outcome iff $x_q=1$, 
i.e., iff $F$ anticommutes with $Z_q$. It keeps $x_q$ but resets $z_q$ to $0$, since it 
leaves $q$ in an eigenstate of $Z_q$. The measurement step is thus linear in $F$'s bits. 
Whether a fault flips a given outcome, in this way or by a readout error, is thus linear 
in the fault. Its pattern is linear in the faults as well, as each detector is a fixed XOR 
of outcomes. Lastly, a conditional Pauli $P_c$ fires on the XOR of some outcomes, making it 
linear in the faults. If a fault flips an odd number of them, whether $P_c$ is applied or not 
changes relative to the fault-free circuit. This multiplies the propagated fault by $P_c$. 
Since every step is $\GF$-linear, the residual and pattern of a higher-order event composed 
of several faults are simply the XOR of those of its constituent faults. The syndrome of 
a higher-order event composes similarly. This is because the product of two Paulis $PQ$ 
anticommutes with a row of $\mathcal R$ iff exactly one of $P$ and $Q$ does, i.e., 
$\text{syn}_{\mathcal R}(PQ)=\text{syn}_{\mathcal R}(P)\oplus\text{syn}_{\mathcal R}(Q)$. 
This bypasses an explicit multi-fault propagation. (b) For state preparation, the 
heralded and wasted events are counted only at order one (cf. Sec.~\ref{sec:funsearch}). 
Since we assume that each gate failure, input error, and measurement error occurs with 
a similar physical probability $\sim p$, each fault of a two-qubit gate weighs $1/15$, 
while those of a one-qubit gate or input error weigh $1/3$, and each measurement error 
weighs $1$. The herald and waste are calculated by performing a weighted sum of the 
faults, using these weights. Thus they approximately equal the probability, in units 
of the physical error rate $p$, of discarding the shot under post-selection. 
(c) For logical gates, the decoder discards a shot only when its bin is ambiguous 
(cf. Sec.~\ref{sec:funsearch}). Faults with the same detector pattern and the same full 
syndrome, i.e., on every row of $\mathcal R$ (the code checks and the logical operators), 
are identical for all purposes. We therefore group them into classes. The full syndrome 
splits into the code-check syndrome, which together with the pattern forms the bin, and 
a logical part. We sort the classes by bin and then by logical part, and do the same for 
the pairs of classes, which form the events of order two. After sorting, the events of 
each bin sit next to each other, so a bin is ambiguous when its events carry more than 
one logical part. An event with an empty pattern is silent if its residual is heavy. An 
event with a non-empty pattern in an ambiguous bin is undecodable if its residual is 
heavy, and forfeited otherwise. For $t=3$, a three-fault event can make a bin of one- and 
two-fault events ambiguous. We check this exactly, so the first- and second-order counts 
stay exact. The undecodable and forfeited events of order three are estimated from 
a random sample of third-order events (three simultaneous faults). The results we quote 
are verified with an exact count at every order.

Lastly, we consider the classification of measurements as deterministic or random for the 
gate design and applications. The classification is based on the Choi state, whereas the 
actual circuit acts on codewords. We need to show that the classification on the Choi state 
matches that on every codeword. We consider the deterministic and random cases separately. 
A measurement is deterministic when the measured operator belongs, up to a sign, to the 
stabilizer group of the circuit at the time of measurement. At any point in the circuit, 
the stabilizers of the Choi state contain the code checks, the $Z$ of each ancilla, and the 
Bell generators $\bar X_j\otimes X_{r_j}$ and $\bar Z_j\otimes Z_{r_j}$, each evolved by 
the preceding operations. Since nothing acts on the reference qubits, the latter become 
$\bar X_j'\otimes X_{r_j}$ and $\bar Z_j'\otimes Z_{r_j}$, where primes denote evolved 
logical operators. Since the measured operator does not act on the reference qubits, it 
cannot contain any Bell generators, and is a product of the evolved code checks and ancilla 
$Z$s only. These stabilize the evolved state of every codeword, so the outcome is deterministic 
on every codeword. For a random measurement, the measured operator anticommutes with 
some stabilizer of the Choi state. If this is an evolved check or ancilla $Z$, which 
stabilizes every codeword, the outcome is then random on every codeword as well. Otherwise 
it commutes with all the evolved checks and ancilla $Z$s and anticommutes only with some 
Bell generator. As it does not act on the reference qubit, it must then anticommute with 
the logical operator in the Bell generator. So it must be some logical operator itself. 
Measuring it would collapse the encoded state, so this cannot happen in a valid circuit.

\section{Verification of dynamic Clifford circuits}
\label{app:frame}

Here we use the results of Ref.~\cite{Kliuchnikov2023} to check the validity of our 
dynamic Clifford circuits. This is not trivial as they contain measurements, as well 
as conditional Paulis accepting these measurement outcomes as their condition, and 
validity must be checked for all possible combinations of measurement outcomes. 

\emph{Symplectic part:--} We start with the verification of the symplectic part. A 
circuit contains both random and deterministic measurements, in general. Deterministic 
measurements have definite and known outcomes, and thus contribute only a single branch. 
However, $n_{\mathrm{rand}}$ random measurements have $2^{n_{\mathrm{rand}}}$ branches corresponding to all possible outcomes.
To verify validity, we must ensure that on each of these exponentially many branches, the 
data qubits reach the desired target state $\ket{\psi_\star}$ (Choi state for gates) and 
the ancillae are disentangled from them. This rules out brute-force enumeration. However, 
on restricting the conditions of the conditional Paulis to XOR/parity of outcomes, we can 
verify validity with only $n_{\mathrm{rand}}+1$ validity checks instead of $2^{n_{\mathrm{rand}}}$. We show this below:

(a) A measurement of some Pauli $Z_q$ is random if $Z_q$ anticommutes with some generator 
of the current state, and deterministic otherwise, as then $\pm Z_q$ belongs to the state's 
stabilizer group. Thus, whether a given measurement is random or deterministic is decided 
solely by the symplectic part (the $X$ and $Z$ bits) of the stabilizer tableau. This is a 
property of the circuit, independent of the measurement outcomes. This is because the branches 
differ only in the signs introduced by random measurements, and which conditional Paulis are 
applied, and Paulis alter only the signs of the tableau. In either case, the symplectic part 
is unchanged.

(b) We restrict to the class of stabilizer circuits considered by Ref.~\cite{Kliuchnikov2023}, 
containing Clifford gates, Pauli (here, $Z$) measurements, and Pauli gates conditioned on the parity 
of measurement outcomes. We define $s_j$ as the outcome of the $j$th random measurement, so 
that a branch is labeled by $s=(s_1,\ldots,s_{n_{\mathrm{rand}}})\in\{0,1\}^{n_{\mathrm{rand}}}$, with $s=0$ the reference branch.
Our circuits start from a fixed stabilizer state $\ket{\psi_{\mathrm{in}}}$, which is $\ket{0\cdots0}$ 
for state preparation, and the input Choi state for gate design (Appendix~\ref{app:gatechoi}). 
On branch $s$, the circuit acts as a linear map $K_s$, where the $s$ 
dependence arises from the random outcomes and the conditional Paulis conditioned on them. 
As in Ref.~\cite{Kliuchnikov2023}, we start from $\ket{0\cdots0}$, and write our $\ket{\psi_{\mathrm{in}}}$ 
as $\ket{\psi_{\mathrm{in}}}=V\ket{0\cdots0}$, where $V$ is a Clifford, which is always 
possible~\cite{AaronsonGottesman2004}. We then apply their results to  
$K_s\ket{\psi_{\mathrm{in}}}=K_s V\ket{0\cdots0}$. Now, according to Ref.~\cite{Kliuchnikov2023}, 
there exist a Clifford $C$ and binary matrices $A$ and $\mu$, such that the final state 
$\ket{\psi(s)}$ and the outcome $o_i(s)$ of the $i$th deterministic measurement satisfy
\begin{subequations}
\begin{align}
  \ket{\psi(s)} &\propto K_s V\ket{0\cdots0} \propto C X^{As}\ket{0\cdots0},
  \label{eq:frame_state0}\\
  o_i(s) &= o_i(0)\oplus\mu_i\cdot s,
  \label{eq:frame_outcome}
\end{align}
\end{subequations}
where the row $\mu_i\in\{0,1\}^{n_{\mathrm{rand}}}$ of $\mu$ specifies how $o_i$ depends on the random outcomes. 
If a measurement of $Z_q$ is deterministic, then $\pm Z_q$ belongs to the stabilizer group (see (a)),
and its outcome is given by this sign. However, this may depend on earlier random outcomes. 
For example, measuring $Z_1$ on $\ket{+}$ gives a random $s_1$ and leaves the state stabilized 
by $(-1)^{s_1}Z_1$. Measuring $Z_1$ again is then deterministic with outcome $s_1$.
Also, for some bit vector $v$, $X^{v}=\prod_q X_q^{v_q}$ applies $X$ to every qubit $q$ for 
which $v_q=1$, so $X^{As}\ket{0\cdots0}=\ket{As}$. At $s=0$, Eq.~\eqref{eq:frame_state0} gives 
the reference state $\ket{\psi(0)}\propto C\ket{0\cdots0}$, which is produced by a single 
simulation. We now express every branch in terms of 
this reference. Using $X^{As}=\prod_j (X^{a_j})^{s_j}$, with $a_j$ the $j$th column of $A$, 
we obtain
\begin{align}
  \ket{\psi(s)} &\propto \bigl(CX^{As}C^\dagger\bigr)\,C\ket{0\cdots0}\nonumber\\
  &\propto\prod_j\bigl(CX^{a_j}C^\dagger\bigr)^{s_j}\,C\ket{0\cdots0} \nonumber\\
  &\propto M(s)\ket{\psi(0)},\qquad M(s)=M_1^{s_1}\cdots M_{n_{\mathrm{rand}}}^{s_{n_{\mathrm{rand}}}},
  \label{eq:frame}
\end{align}
where each $M_j=CX^{a_j}C^\dagger$ is a Pauli, since $C$ is Clifford.

(c) Finally, we show that only $n_{\mathrm{rand}}+1$ checks, the reference branch with all random outcomes
set to zero $\ket{\psi(0)}$, and $n_{\mathrm{rand}}$ branches with a single outcome flipped
$\,M_1\ket{\psi(0)},\,\ldots,\,M_{n_{\mathrm{rand}}}\ket{\psi(0)}$, determine all $2^{n_{\mathrm{rand}}}$ branches. Note that
$\ket{\psi(s)}$ in Eq.~\eqref{eq:frame} is the state defined on all qubits, which includes 
the data, reference, as well as ancilla qubits. The target $\ket{\psi_\star}$ is defined 
on the set $D$ containing the data qubits, and for gates also the reference qubits. The 
ancillae $A$ may end up in any state, as long as they are disentangled from $D$. We start 
by checking that the reference branch reaches the target, i.e., 
$\ket{\psi(0)}=\ket{\psi_\star}\otimes\ket{\phi}_a$, for some ancilla state 
$\ket{\phi}_a$. Every other branch differs from the reference only by the Pauli $M(s)$. 
Since $M_j$ is Pauli for each $j$, it factorizes as $M_j\propto P_j\otimes Q_j$, with Pauli
strings $P_j$ acting on $d$ and $Q_j$ on $a$, so that
\begin{align}
  \ket{\psi(s)}\propto P(s)\ket{\psi_\star}\otimes Q(s)\ket{\phi}_a,
\end{align}
with $P(s)=P_1^{s_1}\cdots P_{n_{\mathrm{rand}}}^{s_{n_{\mathrm{rand}}}}$ and $Q(s)=Q_1^{s_1}\cdots Q_{n_{\mathrm{rand}}}^{s_{n_{\mathrm{rand}}}}$. Thus, the data
$d$ and ancillae $a$ remain disentangled on every branch, with the ancillae in some 
irrelevant state $Q(s)\ket{\phi}_a$. So we only need to check that the state 
$P(s)\ket{\psi_\star}$ on $d$ reaches the target. The branch with the $j$th outcome 
flipped reaches the target iff $P_j\ket{\psi_\star}\propto\ket{\psi_\star}$, which implies 
$P_j\ket{\psi_\star}=\pm\ket{\psi_\star}$ as $P_j^2=I$, i.e., either $P_j$ or $-P_j$ 
belongs to the target's stabilizer group. Provided $P_j\ket{\psi_\star}=\pm \ket{\psi_\star}$ 
for all $j=1,\ldots, n_{\mathrm{rand}}$, it is clear that $P(s)\ket{\psi_\star}=\pm\ket{\psi_\star}$ on every
branch $s$. Hence, it is sufficient to check the reference branch, and the $n_{\mathrm{rand}}$ branches each
with a single outcome flipped. 

\emph{Signs:--} If the symplectic part matches, the branches may differ from the target 
only in the signs of the generators. This is the situation when the conditional Pauli 
corrections can repair the circuit and ensure that the signs match in every branch. 

Before proceeding, we show a result that the signs combine $\GF$-linearly. To obtain the 
signs, we simply operate on the state with the generator. Consider an unsigned generator 
$g$, and start with the reference branch. Applying $g$ to the reference state yields 
$g|\psi(0)\rangle=(-1)^{r_g(0)}|\psi(0)\rangle$, implying that the signed generator $g$ 
should have the sign $r_g(0)$ in the reference branch. To establish the signs of the other 
branches, use the branch-specific state obtained above in Eq.~\eqref{eq:frame}, 
\begin{align}
  gM(s)|\psi(0)\rangle =& (-1)^{\langle M(s),g \rangle }M(s)g|\psi(0)\rangle\nonumber\\
  =& (-1)^{r_g(0) \oplus \langle M(s),g \rangle }M(s)|\psi(0)\rangle,
\end{align}
where $\langle P,Q\rangle\in\GF$ is the symplectic product of two Pauli strings, defined by
$PQ=(-1)^{\langle P,Q\rangle}QP$. In terms of their $X$ and $Z$ bits,
$\langle P,Q\rangle=x_P\cdot z_Q\oplus z_P\cdot x_Q$, which is linear in each argument, so
$\langle M(s),g\rangle=\bigoplus_j s_j\langle M_j,g\rangle$.
Now, we get the sign
\begin{align}
  r_g(s) = &  r_g(0) \oplus \langle M(s),g \rangle\nonumber\\
   =& r_g(0) \oplus V_g \cdot s;\quad V_{g,j}=\langle M_j,g \rangle
\end{align}
Here $V_{g,j}$ is simply the sign flip, relative to the reference branch's sign $r_g(0)$, 
introduced in the branch $j$. Now, the signs in each branch must match that of the target 
for that generator ($r_{g,\star}$), which yields
\begin{align}
  r_g(0) \oplus V_g \cdot s &= r_{g,\star} \implies V_g \cdot s = r_g(0)\oplus r_{g,\star}.
\end{align}
Considering all the generators, we obtain the $\GF$-linear system
\begin{align}
  V\cdot s =&  r(0)\oplus r_\star.
\end{align}

Now we look at a quantity that gives the fraction of branches whose signs match the target's 
signs
\begin{equation}
  \text{branch fraction} =
  \begin{cases}
    2^{-\mathrm{rank}(V)}, & Vs=r(0)\oplus r_\star \text{ solvable},\\[2pt]
    0, & \text{otherwise}.
  \end{cases}
  \label{eq:branchfrac}
\end{equation}

\emph{Match:--} An invalid circuit must still be scored, as the search benefits from a 
``gradient'' towards valid circuits. This involves quantifying how close the symplectic 
part of the tableau, and the signs, are to those of the target state. For the latter, we 
already have Eq.~\eqref{eq:branchfrac}. For the former, we calculate the similarity of 
the tableaux of the final state and the target. Let $b$ and $b_\star$ be the flattened
bit strings formed by the $X$, $Z$ and sign bits of the canonical tableaux of a branch and of 
the target. We define tableau similarity as their Jaccard similarity $|b\wedge b_\star|/|b\vee b_\star|$, 
where $|\cdot|$ counts the ones, averaged over the $n_{\mathrm{rand}}+1$ simulated branches. The match (cf. 
Eq.~\eqref{eq:score}) is
\begin{equation}
  \text{match}=\tfrac{1}{2}\big(\text{tableau similarity}+\text{branch fraction}\big).
\end{equation}
It is $1$ for a valid circuit and is smaller otherwise.

\section{Pseudocode}
\label{app:pseudo}

In this section we provide pseudocode for a concise description of our workflow.

\subsection{Agentic FunSearch loop}

\begin{flushleft}\footnotesize
\texttt{tools} $\leftarrow$ \textsc{Test}, \textsc{Python}\quad\textcolor{gray}{\textit{// see below}}\\
\textbf{for} \texttt{generation}$\ =1,\ \ldots,\ N_\text{gen}$\\
\quad \textbf{for} \texttt{island}$\ =1,\ \ldots,\ N_\text{islands}$\\
\quad\quad \texttt{parents} $\leftarrow$ best of \texttt{island} and $K-1$ of its other programs\\
\quad\quad \texttt{prompt} $\leftarrow$ task, circuit class API, \texttt{parents} with\\
\quad\quad\quad feedback, scratchpad notes of \texttt{island}\\
\quad\quad (\texttt{program}, \texttt{notes}) $\leftarrow$ \textsc{LLM}(\texttt{prompt}, \texttt{tools})\\
\quad\quad \textbf{for} \texttt{size} in \texttt{grid}\\
\quad\quad\quad \texttt{circuit} $\leftarrow$ \texttt{program}(\texttt{size})\\
\quad\quad\quad $\sigma_\texttt{size}\leftarrow$ score of \textsc{Verify}(\texttt{circuit})\\
\quad\quad insert \texttt{program} into \texttt{island}, in the cluster of\\
\quad\quad\quad $\sigma_\texttt{size}$ at the largest size\\
\quad every \texttt{migrate} generations copy the best program\\
\quad\quad of every \texttt{island} to \texttt{island}$+1$\\
\textbf{return} the best program, ranked by $\sigma_\texttt{size}$ at all sizes
\end{flushleft}
\noindent The agentic FunSearch loop, as described in Sec.~\ref{sec:funsearch}. The run takes place 
on $N_\text{islands}$ independent islands, and covers $N_\text{gen}$ generations. A \texttt{size} 
denotes a single tuple $(\ndata,\nanc)$ passed to the program to build the layout, e.g., the surface
code layout at a single distance. The \texttt{grid} is the set of sizes at which the program is scored. 
Each island maintains its own database of programs clustered by their score $\sigma$, the lexicographic 
tuple defined in Eq.~\eqref{eq:score} (Eq.~\eqref{eq:score_gates} for gates), at each \texttt{size}. 
The LLM can call two tools before returning a program: \textsc{Test} accepts only a program, scores it 
on \texttt{grid}, and returns the verifier feedback that a parent carries in the prompt. \textsc{Python} 
runs any Python code with the circuit class, the targets and the verifier in scope, and returns what 
the code prints, so that, e.g., many variants can be compared in one call, or single faults traced by 
hand. Along with the program, the LLM returns scratchpad notes, shown in the island's next prompt. 
The clusters are
keyed by the score at the largest size, and the programs within a cluster are ranked by the
scores at the smaller sizes, which gives the lexicographic order of Sec.~\ref{sec:funsearch}.
Every \texttt{migrate} generations, the best program of each island is copied to the next island.

\subsection{Verifier}

\begin{flushleft}\footnotesize
\textsc{TableauSimulate}(\texttt{circuit}, \texttt{force\_rand}):\\
\quad run \texttt{circuit} on the input tableau with the $j$th random\\
\quad outcome set to \texttt{force\_rand}$_j$\\
\quad \textbf{return} final tableau $\tau$, outcomes $o$, number $n_{\mathrm{rand}}$\\
\quad of random outcomes\\[4pt]
\textsc{ScoreFT}(\texttt{circuit}, \texttt{detectors}):\\
\quad initialize \texttt{faults}\\
\quad propagate \texttt{faults} to get their residual $E_f$ on data qubits\\
\quad\quad and its pattern $p_f$ of flipped \texttt{detectors}\\
\quad $H\leftarrow\{f\in \texttt{faults}:\wtS(E_f)>\min(1,t)\}$\quad\textcolor{gray}{\textit{// heavy}}\\
\quad silent$_1\leftarrow|\{f\in H:p_f=\emptyset\}|$\\
\quad herald $\leftarrow$ weight of $\{f\in H:p_f\neq\emptyset\}$\\
\quad waste $\leftarrow$ weight of $f\notin H$ with $p_f\in\{p_h:h\in H,\ p_h\neq\emptyset\}$\\
\quad \textbf{for} $i=2,\dots,t$ \textbf{do}\\
\quad\quad silent$_i\leftarrow$ number of sets of $i$ faults at distinct\\
\quad\quad\quad locations with $\oplus_f\,p_f=\emptyset$ and $\wtS(\prod_f E_f)>i$\\
\quad \textbf{return} silent$_1$,\ldots,silent$_t$, herald, waste
\end{flushleft}
\noindent Here, we show the two subroutines used by the verifier below. \textsc{TableauSimulate} 
runs the circuit \texttt{circuit} on the input stabilizer tableau, with the $j$th random
measurement outcome forced to \texttt{force\_rand}$_j$. \textsc{ScoreFT} obtains the fault 
tolerance metrics of a valid circuit with detectors \texttt{detectors} (Appendix~\ref{app:faulttol}), 
shown here for state preparation. Each fault in \texttt{faults} is propagated to the end 
of the circuit to obtain its residual $E_f$ on the data qubits, and then classified as 
described in Appendix~\ref{app:faulttol}. For gates, the faults are instead binned by their 
detector pattern and code-check syndrome, and the undecodable and forfeited events of 
Eq.~\eqref{eq:score_gates} are counted. This follows similarly and is omitted here for 
brevity. With the CSS simplifications of Appendix~\ref{app:faulttol}, $\wtS$ is evaluated
only on the $X$ ($Z$) part for $\ket{0_L}$ ($\ket{+_L}$), and separately on the $X$ and $Z$
parts for gates.

\begin{flushleft}\footnotesize
\textsc{Verify}(\texttt{circuit}):\\
\quad $(\tau_0,o^{(0)},n_{\mathrm{rand}})\leftarrow$ \textsc{TableauSimulate}(\texttt{circuit}, $0$)\\
\quad \textbf{for} $j=1,\dots,n_{\mathrm{rand}}$\\
\quad\quad $(\tau_j,o^{(j)})\leftarrow$ \textsc{TableauSimulate}(\texttt{circuit}, $e_j$)\\
\quad \texttt{detectors} $\leftarrow$ (1) deterministic outcomes,\\
\quad\quad (2) fixed XORs, e.g., a random outcome and its repeat\\
\quad \textbf{if} $\tau_j=\tau_\star\ \forall j=0,\dots,n_{\mathrm{rand}}$ \quad\textcolor{gray}{\textit{// valid}}\\
\quad\quad \textbf{return} match 1 and \textsc{ScoreFT}(\texttt{circuit}, \texttt{detectors})\\
\quad $V_{:,j}\leftarrow r(\tau_j)\oplus r(\tau_0)$, \ $v\leftarrow r(\tau_0)\oplus r_\star$\\
\quad \textbf{if} $\tau_0\neq\tau_\star$ up to signs\quad\textcolor{gray}{\textit{// wrong Pauli part}}\\
\quad\quad branch\_frac $\leftarrow 0$\\
\quad \textbf{else if} $Vs=v$ is solvable\\
\quad\quad branch\_frac $\leftarrow 2^{-{\rm rank}\,V}$ \\
\quad \textbf{else}\\
\quad \quad branch\_frac $\leftarrow 0$\\
\quad sim $\leftarrow$ mean Jaccard similarity of $\tau_j$ to $\tau_\star$\\
\quad \textbf{return} match $(\text{branch\_frac}+\text{sim})/2$
\end{flushleft}
\noindent The verification routine starts with tableau simulations using 
\textsc{TableauSimulate} to identify the detectors. A valid circuit, whose final tableau matches 
the target on every branch, is scored by \textsc{ScoreFT}, and an invalid one by its match.

\subsection{Prompt}
\label{app:prompt}

\begin{flushleft}\footnotesize
\textsc{Prompt}(\texttt{island}, \texttt{generation}):\\
\quad\textcolor{gray}{\textit{// fixed}}\\
\quad role: write a program returning a fault-tolerant Clifford circuit\\
\quad target: the task in words, the initial and target generators\\
\quad primer: the Choi construction, how gates and measurements\\
\quad\quad act on the tableau, the fault locations and the repair criterion\\
\quad layout: data and ancilla qubits, couplings, evaluated sizes, $t$\\
\quad signature for circuit program: \texttt{build\_circuit(n\_data, n\_anc)}\\
\quad\quad with \texttt{CONCEPT} docstring describing the circuit.\\
\quad rules: design protocol (ranking, how to improve), circuit API,\\
\quad\quad gate set, feedback format, tool budgets, answer format\\
\quad\textcolor{gray}{\textit{// changes every generation}}\\
\quad $K$ parent programs from \texttt{island}, with score $\sigma$ and feedback:\\
\quad\quad per size: validity, event counts, depth, gate counts, match\\
\quad\quad packed circuit; which measurements are detectors\\
\quad\quad silent events: the gates they originate at, residuals\\
\quad\quad ambiguous bins: candidate fixes, their gates, residuals\\
\quad recent errors, notes of \texttt{island}, search progress\\
\quad task: write an improved program that differs from every parent
\end{flushleft}
\noindent The prompt. Its fixed part comes first for prompt caching.

\subsection{LLM-generated circuits}
\label{app:llmprogram}

Here we present four programs written by the LLM, and two of their circuits.
In each case, the LLM writes a concise description of the design principle in 
the docstring \texttt{CONCEPT}. The prompt asks the LLM to state how the freshly 
generated circuit changes relative to the parent circuits from the previous 
generations. So the description in all four programs below is relative to the
parents rather than self-contained. This was our design choice. To avoid wasting
generations, each new program must differ from its parents, and we expect that 
explicitly instructing the LLM to name the change would encourage it to produce a 
distinct program. Together with the docstrings of all the programs in the database, 
as well as the scratchpad notes, this lets one reconstruct the full design logic. 
Nevertheless, this instruction in the prompt can simply be replaced with the user's 
requirements, e.g., a request for a self-contained description.

\subsubsection{Rotated surface code, multi-size}
\label{app:llmprogram_surface}

The following program prepares $\ket{0_L}$ of the rotated surface code on the native 
layout (``multi-size'' in Tab.~\ref{tab:states}). This program constructs the 
circuits for all distances, and we have tested it for $d\in\{3,5,7,9\}$. It provides 
a generalized solution for all $d\geq 5$, and treats $d=3$ ($\ndata=9$) in a separate branch.

\begin{lstlisting}[frame=tb, framesep=3pt, columns=flexible, breakautoindent=true, breakindent=0pt, postbreak=\mbox{$\hookrightarrow$\ }]
def build_circuit(n_data, n_anc):
    """CONCEPT: Keep the dual-basis encoder and the
    geometrically pruned deterministic checks, but change
    the feedforward decoder: grow each X-check correction
    string toward the opposite incident check by reversing
    the incidence-tree tie breaks. This changes only the
    outcome-to-Pauli map, not the measurement protocol or
    detector set.
    """
    qc = Circuit(n_data, n_anc)
    d = int(math.sqrt(n_data))
    index = {q: i for i, q in enumerate(qc.data)}
    xchecks, zchecks = [], []
    for a in qc.anc:
        support = sorted(qc.neighbors(a), key=lambda q: index[q])
        positions = [divmod(index[q], d) for q in support]
        if len(support) == 4:
            is_x = (min(r for r, c in positions) + min(c for r, c in positions)) % 2 == 0
            corners = [0, 1, 2, 3]
        else:
            is_x = positions[0][0] == positions[1][0]
            if is_x:
                corners = [2, 3] if positions[0][0] == 0 else [0, 1]
            else:
                corners = [1, 3] if positions[0][1] == 0 else [0, 2]
        order = [1, 0, 3, 2] if is_x else [1, 3, 0, 2]
        slots = [None] * 4
        for q, corner in zip(support, corners):
            slots[order.index(corner)] = q
        if not is_x and len(support) == 4 and min(r for r, c in positions) == 0:
            slots[0] = None
        (xchecks if is_x else zchecks).append((a, slots, support))

    if n_data == 9:
        for a, slots, support in xchecks:
            qc.h(a)
        for step in range(4):
            for a, slots, support in xchecks:
                if slots[step] is not None:
                    qc.cnot(a, slots[step])
            for a, slots, support in zchecks:
                if slots[step] is not None:
                    qc.cnot(slots[step], a)
        for a, slots, support in xchecks:
            pivot = next(q for q in support if sum(q in s for _, _, s in xchecks) == 1)
            qc.cnot(pivot, a)
        for a, slots, support in zchecks:
            qc.measure(a)
        return qc

    paths = {}
    for radius in range(d + 1):
        for j, (a, slots, support) in enumerate(xchecks):
            if j in paths:
                continue
            for q in reversed(support):
                other = [k for k, (_, _, s) in enumerate(xchecks) if k != j and q in s]
                if not other:
                    paths[j] = [q]
                    break
                if other[-1] in paths and len(paths[other[-1]]) <= radius:
                    paths[j] = [q] + paths[other[-1]]
                    break

    for q in qc.data:
        qc.h(q)
    for a, slots, support in zchecks:
        qc.h(a)
    for step in range(4):
        for a, slots, support in xchecks:
            if slots[step] is not None:
                qc.cnot(slots[step], a)
        for a, slots, support in zchecks:
            if slots[step] is not None:
                qc.cnot(a, slots[step])
    outcomes = [qc.measure(a) for a, slots, support in xchecks]
    for q in qc.data:
        qc.h(q)
    for a, slots, support in zchecks:
        qc.h(a)
    for a, slots, support in zchecks:
        qc.measure(a)
    for q in qc.data:
        controls = [outcomes[j] for j in range(len(xchecks)) if q in paths[j]]
        if controls:
            qc.z(q, cond=controls)
    return qc
\end{lstlisting}

\begin{figure}[t]
\centering
\includegraphics[width=0.975\columnwidth]{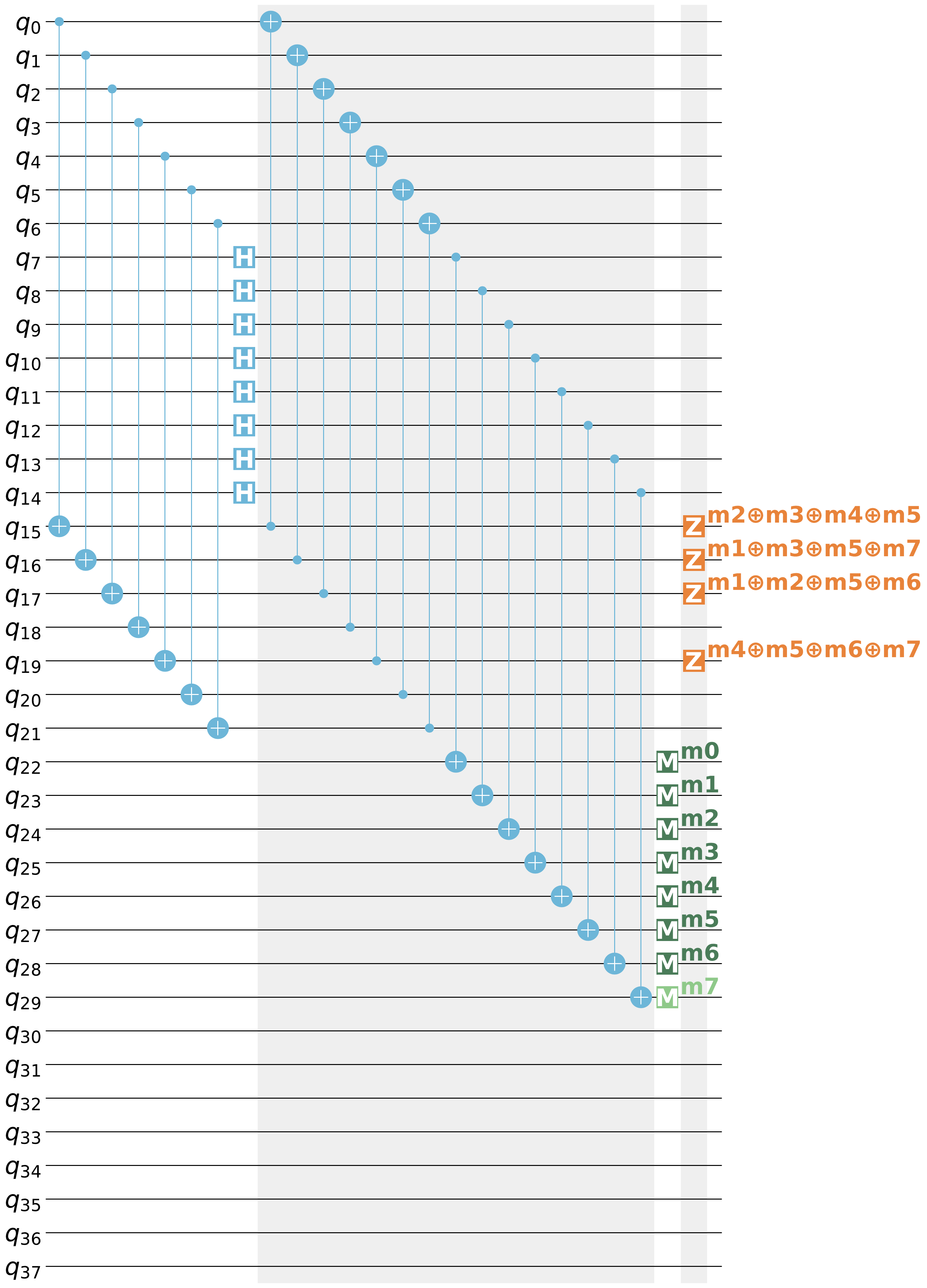}
\caption{The Reed-Muller (RM) to Steane (S) switch with all-to-all connectivity. It reads 
$q_7$-$q_{14}$ of the RM block $q_0$-$q_{14}$ in the $X$ basis, and moves the face 
$q_0$-$q_6$ to the fresh S block $q_{15}$-$q_{21}$. $q_{22}$-$q_{37}$ are ancillae. Dark 
green: random measurements; light green: here $m_7$, which equals $m_0\oplus\cdots\oplus m_6$ 
without faults, so $m_0\oplus\cdots m_6\oplus m_7$ is a detector. The circuit docstring 
mentions it; orange: conditional Paulis, with their conditions beside them. Alternate 
layers are shaded grey.
}
\label{fig:rmtos_circuit}
\end{figure}

\subsubsection{Reed-Muller to Steane switch}
\label{app:llmprogram_rmtos}

\begin{figure*}[!htbp]
\centering
\includegraphics[width=0.975\textwidth]{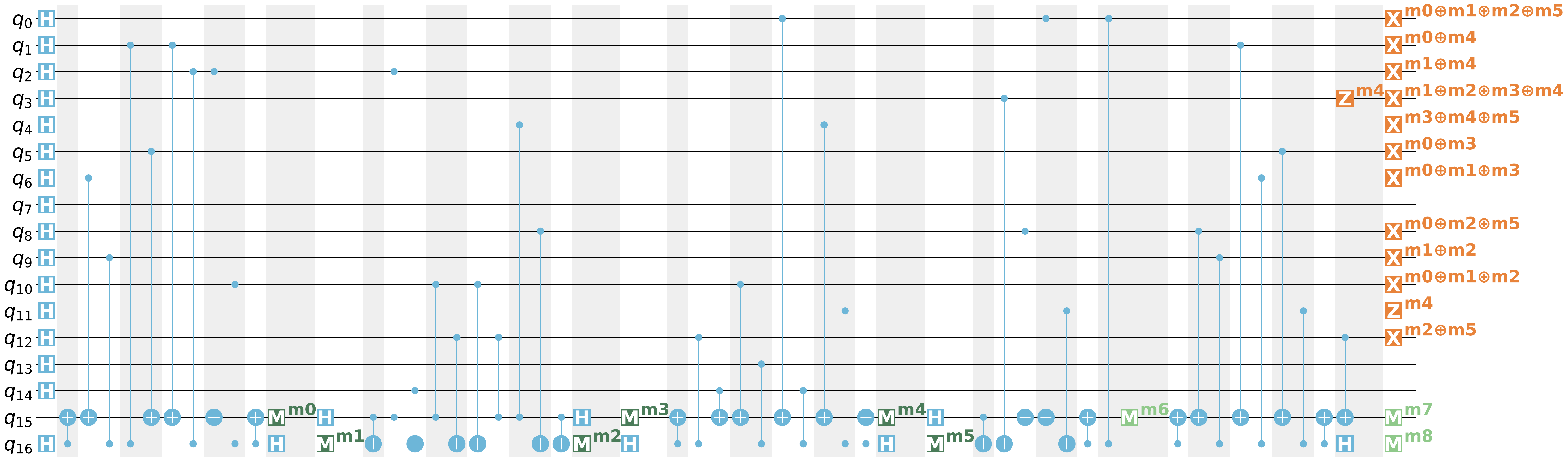}
\caption{The logical Hadamard on the $[[15,1,3]]$ code with all-to-all connectivity and two 
ancillae. $q_0$-$q_{14}$ are the data qubits, and $q_{15}$, $q_{16}$ the ancillae. Colours 
and shading are as in Fig.~\ref{fig:rmtos_circuit}.
}
\label{fig:rmh_circuit}
\end{figure*}

The following program implements the Reed-Muller to Steane switch 
(Tab.~\ref{tab:applications}), with its circuit in Fig.~\ref{fig:rmtos_circuit}. 

\begin{lstlisting}[frame=tb, framesep=3pt]
def build_circuit(n_data, n_anc):
    """CONCEPT: Change the gate body to parallel X-basis
    puncturing: measure Reed-Muller data 7--14 while
    retaining their known even outcome parity as a detector.
    Transfer the remaining Steane block into data 15--21
    with two CNOTs per wire, then restore its X-check signs
    with even-weight Z feedforward. Implement each X readout
    by rotating the discarded data, saving a layer over
    ancilla-controlled readout. The parity detector and
    output check syndrome distinguish readout faults without
    postselection.
    """
    qc = Circuit(n_data, n_anc)

    outcomes = []
    for j in range(8):
        qc.h(qc.data[7 + j])
        qc.cnot(qc.data[7 + j], qc.anc[j])
        outcomes.append(qc.measure(qc.anc[j]))

    for j in range(7):
        qc.cnot(qc.data[j], qc.data[15 + j])
        qc.cnot(qc.data[15 + j], qc.data[j])

    corrections = (
        (16, (1, 3, 5, 7)),
        (17, (1, 2, 5, 6)),
        (15, (2, 3, 4, 5)),
        (19, (4, 5, 6, 7)),
    )
    for q, bits in corrections:
        qc.z(qc.data[q], cond=[outcomes[j] for j in bits])

    return qc
\end{lstlisting}

\subsubsection{Logical Hadamard on the $[[15,1,3]]$ code}
\label{app:llmprogram_rmh}

The following program implements the logical Hadamard on the $[[15,1,3]]$ code with two 
ancillae (Tab.~\ref{tab:gates}), with its circuit in Fig.~\ref{fig:rmh_circuit}.

\begin{lstlisting}[frame=tb, framesep=3pt]
def build_circuit(n_data, n_anc):
    """CONCEPT: Change the tail detectors and their Bell
    decoder: replace the last Z probe {3,6,8,9} by its
    product with the first tail check, {0,6,9,11}. Reverse
    the first tail's decoding CNOT, exchange the
    continuing ancilla roles, and reorder its routed
    couplings so the complementary phase frame is on
    {3,11}. Keep three alternating, mutually flagged gauge
    pairs after transversal H and defer XOR frames.
    """
    qc = Circuit(n_data, n_anc)
    gauges = [[6, 5, 1, 2], [9, 1, 2, 10],
              [14, 12, 10, 8], [2, 10, 12, 4],
              [14, 10, 0, 4], [12, 13, 14, 11]]
    fixes = [[0, 2, 8, 10], [0, 2, 4, 6],
             [3, 4, 9, 10], [1, 2, 5, 6],
             [1, 2, 3, 4], [0, 4, 8, 12]]
    pending = []
    previous = [[], []]

    for q in qc.data:
        qc.h(q)

    for k in range(3):
        z = k % 2
        x = 1 - z
        a, b = qc.anc[z], qc.anc[x]
        qc.h(b)
        qc.cnot(b, a)
        for i, j in zip(gauges[2*k], gauges[2*k+1]):
            qc.cnot(qc.data[i], a)
            qc.cz(b, qc.data[j])
        qc.cnot(b, a)
        ma = qc.measure(a)
        qc.h(b)
        mb = qc.measure(b)
        pending.append((fixes[2*k], [ma] + previous[z]))
        pending.append((fixes[2*k+1], [mb] + previous[x]))
        previous[z] = [ma]
        previous[x] = [mb]

    b, a = qc.anc[0], qc.anc[1]
    qc.h(b)
    qc.cnot(b, a)
    for i, j in zip([8, 0], [3, 11]):
        qc.cnot(qc.data[i], b)
        qc.cnot(qc.data[j], a)
    qc.cnot(a, b)
    qc.measure(b)
    a, b = b, a

    qc.cz(b, qc.data[0])
    qc.cnot(b, a)
    for i, j in zip([8, 1, 5], [9, 6, 11]):
        qc.cnot(qc.data[i], a)
        qc.cz(b, qc.data[j])
    qc.cnot(b, a)
    qc.cnot(qc.data[12], a)
    qc.measure(a)
    qc.h(b)
    qc.measure(b)

    for i in [3, 11]:
        qc.z(qc.data[i], cond=previous[0])

    for q in range(n_data):
        labels = set()
        for row, cond in pending:
            if q in row:
                labels ^= set(cond)
        if labels:
            qc.x(qc.data[q], cond=sorted(labels))
    return qc
\end{lstlisting}

\subsubsection{Preparing $\ket{010101000000}_L$ on the $[[144,12,12]]$ gross code }
\label{app:llmprogram_gross010101000000}

The following program prepares $\ket{010101000000}_L$ on the $[[144,12,12]]$ gross code 
with $72$ ancillae (Tab.~\ref{tab:states}).

\begin{lstlisting}[frame=tb, framesep=3pt]
def build_circuit(n_data, n_anc):
    """CONCEPT: Project sparse X checks in the rotated basis, then verify all 72 Z checks using the projection ancillas without resetting them. Interleave two fresh detectors with the main Z pass so their readouts finish one layer earlier, without changing the order of syndrome couplings on any data wire. Move the left phase frame past the verification couplings to free the early slots.
    """
    qc = Circuit(n_data, n_anc)
    h = n_data // 2
    xs, zs = [], []
    for p in range(h):
        i, j = divmod(p, 6)

        def c(di, dj):
            return 6 * ((i + di) % 12) + (j + dj) % 6

        xs.append((c(3, 0), c(0, 1), c(0, 2),
                   h + c(0, 3), h + c(1, 0), h + c(2, 0)))
        zs.append((c(0, -3), c(-1, 0), c(-2, 0),
                   h + c(-3, 0), h + c(0, -1), h + c(0, -2)))

    active = list(range(64)) + [66, 67]
    checks = list(range(h))
    early = (68, 69)
    rest = [p for p in checks if p not in early]

    # Carry measurement-sign masks through elimination and back-substitution.
    basis = {}
    for p in active:
        row = sum(1 << v for v in xs[p])
        mask = 1 << p
        while row:
            v = (row & -row).bit_length() - 1
            if v not in basis:
                basis[v] = (row, mask)
                break
            old, signs = basis[v]
            row ^= old
            mask ^= signs

    phase = {}
    for v in sorted(basis, reverse=True):
        row, mask = basis[v]
        for w in phase:
            if (row >> w) & 1:
                mask ^= phase[w]
        phase[v] = mask

    def couple(rows, ps, leg):
        for p in ps:
            qc.cnot(qc.data[rows[p][leg]], qc.anc[p])

    for q in qc.data:
        qc.h(q)
    for leg in (1, 2, 0, 4, 5, 3):
        couple(xs, active, leg)
    m = {p: qc.measure(qc.anc[p]) for p in active}
    for q in qc.data:
        qc.h(q)

    # Support of Xbar_2 Xbar_4 Xbar_6.
    flip = (
        3, 10, 17, 19, 25, 28, 34, 39, 43, 53, 58, 67,
        77, 80, 82, 85, 87, 96, 100, 101, 104, 113, 116, 118,
        120, 121, 132, 136, 137, 143,
    )

    def frame(left):
        for v, mask in phase.items():
            if (v < h) == left:
                qc.z(qc.data[v], cond=[m[p] for p in active if (mask >> p) & 1])

    couple(zs, checks, 4)
    for v in flip:
        if v >= h:
            qc.x(qc.data[v])
    frame(False)

    couple(zs, checks, 0)
    couple(zs, early, 3)
    couple(zs, checks, 1)
    couple(zs, early, 2)
    couple(zs, rest, 3)
    couple(zs, early, 5)
    for p in early:
        qc.measure(qc.anc[p])

    # Cancel each retained projection bit at its final detector coupling,
    # then undo the temporary change on the data wire.
    for p in active:
        qc.x(qc.data[zs[p][5]], cond=m[p])
    couple(zs, rest, 2)
    couple(zs, rest, 5)
    for p in active:
        qc.x(qc.data[zs[p][5]], cond=m[p])

    for v in flip:
        if v < h:
            qc.x(qc.data[v])
    frame(True)
    for p in rest:
        qc.measure(qc.anc[p])
    return qc
\end{lstlisting}

\subsection{LLM scratchpad notes}
\label{app:scratchpadnotes}
Figs.~\ref{fig:scratchpad_notools} and~\ref{fig:scratchpad} show scratchpad notes from 
the Steane to Reed-Muller switch. Fig.~\ref{fig:scratchpad_notools} is from a search without 
tools (Fig.~\ref{fig:storm_trajectory}), and Fig.~\ref{fig:scratchpad} is from the agentic 
search of Tab.~\ref{tab:applications} and Fig.~\ref{fig:recordratewordcount}, 
with one note from each LLM tier.

\begin{figure}[htb]
\centering
\begin{lstlisting}[language={}, frame=tb, framesep=3pt,
                   basicstyle=\ttfamily\scriptsize, breaklines=true,
                   title={GPT-6-luna [max], no tools, generation 22, island 1}]
### SCRATCHPAD

- **Verified reference:** gen21 is valid with `silent1=0`, `herald_cost=0.87`, `waste_cost=0.40`, and depth 106. Its two reported ambiguous cells both involve detector `m11`.
- **Other parent approaches:** The direct transfer was valid but had 322 silent single faults; the row-serpentine synthesis was valid but had 1090. Neither is used here.
- **Candidate:** Conjugate gen21's full temporal-detector schedule by a different coordinate permutation. For y = bitreverse(i+1), swap the coordinate bits of weights 2 and 8, giving `perm = [3,1,5,0,4,2,6,7,11,9,13,8,12,10,14]`. This preserves the first-seven-qubit Steane code, the Reed-Muller code, and the full-support logical operators. It differs from gen21's weight-2/weight-4 swap.
- The detector schedule and feedforward are unchanged. The candidate is not independently simulator-verified; the coordinate-automorphism conjugation predicts the same valid branches and fault metrics as gen21, not an improvement. Keep it as a distinct, symmetry-based candidate.
- **Deferred:** Find a temporal detector that splits the `m11` ambiguous cells; without verification, adding detector gates could worsen the fault costs.
\end{lstlisting}
\caption{A scratchpad note from a Steane to Reed-Muller switch without tools
(the search in Fig.~\ref{fig:storm_trajectory} ending with non-zero discards), by the 
strong tier (GPT-6-luna [max]). The LLM defers finding a detector, as without any verification 
it risked worsening the fault tolerance metrics. Instead, it submits the parent circuit 
(provided in its prompt) relabeled by a code automorphism, which scores identically. This 
note uses the verifier's internal names: \texttt{herald\_cost} and \texttt{waste\_cost} 
denoting $\text{undecodable}_1$ and $\text{forfeited}_1$, respectively, and ``cells'' 
denoting bins. }
\label{fig:scratchpad_notools}
\end{figure}

\begin{figure*}[p]
\centering
\begin{minipage}[t]{0.49\textwidth}
\begin{lstlisting}[language={}, frame=tb, framesep=3pt,
                   basicstyle=\ttfamily\scriptsize, breaklines=true,
                   title={(a) GPT-6-luna [high], generation 27, island 0}]
### SCRATCHPAD

- **Scope:** Exact verification at `n_data=15`, `n_anc=2`, residual tolerance `t=1`; the certificate is for single faults.
- **KEPT -- new depth-20 body:** In the compressed first Z-check parity, use the opposite temporary-data orientation: `d6->d7; d7->s; d6->d7`, followed by `d9->s`. Keep the deterministic four-probe schedule and the remaining conversion body.
- **Exact test result:** Valid on every branch; `silent1=0`, `herald1=0`, `waste1=0`, depth **20**, **49 operations**, **4 measurements**, **670 single faults**. All measurements are deterministic zero; no feedforward or postselection. The tested circuit is the one submitted below.
- **FAILED orientation:** `d7->d6; d6->s; d7->d6; d9->s` is valid at depth 20 but leaves one silent fault, born at `CNOT d6->a0`, with residual `X{6} Z{9}`. Reversing the temporary-data orientation removes it in the exact test.
- **Reference point:** The prior depth-21 construction is also fully fault tolerant. The newly tested circuit changes the gate body, so it differs from the listed parents.
- **Deferred:** Search for a depth below 20 while retaining zero silent, herald, and waste costs. No claim of improvement beyond the tested single-fault certificate.
\end{lstlisting}
\vspace{6pt}
\begin{lstlisting}[language={}, frame=tb, framesep=3pt,
                   basicstyle=\ttfamily\scriptsize, breaklines=true,
                   title={(b) GPT-6-luna [max], generation 10, island 0}]
### SCRATCHPAD

- **Kept -- shortened flagged-probe construction.** The tested circuit removes the early `CNOT(s,f)` echo pair, reads the four flagged parity probes in ascending order, omits the readout after the first probe, and cancels the adjacent shared `d13 -> s` couplings across that unmeasured boundary. `test_build_circuit` verified it at the evaluated size: valid, `silent1=0`, `herald1=0`, `waste1=0`, depth 38, 59 operations, 5 deterministic measurements, and branch fraction 1. No claim about fault orders above 1.
- **Sweep evidence.** With the early echo pair removed, a sweep of probe orders and omitted-readout positions found 12 zero-cost variants; the best depth was 38. The selected ascending-order variant was then directly verified.
- **Ablations.** Keeping the early echo pair gave a valid zero-cost depth-39 version; removing it yielded the verified depth-38 version. Removing either source-flag or individual probe-flag pairs introduced silent single faults. In one probe-0-flag ablation, the two silent faults were born at `CNOT d11->a0` and `CNOT d14->a0`, with residual `Z{9,10}`; restoring the flag pair eliminated them.
- **Design change.** This differs from the listed parents in both the gate body (the echo pair is removed) and the detector schedule (ascending probe order and a combined first-probe boundary). Retain the flags; do not infer tolerance beyond the verified single-fault order.
\end{lstlisting}
\end{minipage}\hfill
\begin{minipage}[t]{0.49\textwidth}
\begin{lstlisting}[language={}, frame=tb, framesep=3pt,
                   basicstyle=\ttfamily\scriptsize, breaklines=true,
                   title={(c) GPT-6.1-sol [high], generation 12, island 0}]
### SCRATCHPAD

- **Kept -- parallel mutually flagged final probes, depth 26.** Both final exact tests at `(15,2)` certify validity, `silent1=0`, `herald1=0`, `waste1=0`, depth 26, 54 operations, 4 deterministic measurements, 721 faults, branch fraction 1. Certification covers single faults only.
- **Change from the depth-31 parent:** retain the conversion body and merged first detector. Read the source flag immediately afterward, then reuse both ancillae for two parallel Z-check probes.
- **Final probe construction:** prepare `s` in the X basis; `f` remains in the Z basis after its readout. Open with `CNOT(s,f)`. Interleave `CZ(data[a],s)` on `[12,14,10,8]` with `CNOT(data[b],f)` on `[13,11,9,7]`. Close with `CNOT(s,f)`, then measure `s` in X and `f` in Z. The shared CNOTs mutually flag the probes.
- **Kept scheduling details:** retain root order `13,8,10,9,11,12`, cat-check order `9,10,11,13,12,8,14,7`, delayed `d5 -> d13`, and source-flag closure after the first probe's second flag interaction.
- **Verified intermediate -- symmetric CZ probes:** separately reading the old flag, then using two X-basis ancillae and two mutual CZs, gave zero costs at depth 27. Changing one probe to a Z-basis CNOT target removes two Hadamards and gives depth 26.
- **Failed -- merge the old flag readout into the parallel probes:** valid depth-26 variants had silent faults. The tested outermost mutual-CZ placement had 13 silent faults. Reading the old flag separately removed them.
- **Ablations:** deleting any single operation from the final depth-26 circuit either invalidated it or introduced silent faults. Deleting the separate source-flag measurement introduced 13 silent faults. Deleting either shared final CNOT introduced 137 silent faults.
- **Mutual-flag placement:** in the symmetric implementation, opening before all data couplings and closing after all data couplings gave zero costs. Every other tested pair of symmetric boundaries introduced silent faults.
- **Source-flag closure sweep:** retained placement gives zero costs at depth 26. Moving it earlier within the first probe introduced discard costs or silent faults; other zero-cost tested placements reached depth 28 or tied depth 26.
- **Failed -- compress the cat checker through `d13 -> d14`:** replacing its two endpoint couplings by an earlier single coupling introduced 9-10 silent faults in the serial-detector construction.
- **Failed -- compensated compressed branch:** copying `d5` into both `d13` and `d14` permits a shorter checker algebraically, but tested schedules had 4-8 silent faults and nonzero herald cost.
- **Failed -- first parity includes d7 to absorb a flag interaction:** swept three supports containing d7, their permutations, flag boundaries, and source-flag closing positions. Best tested variants still had 24 silent faults. Mixed-Pauli cancellations depend on the cat preparation and merged-check support.
- **Deferred:** redesigning the conversion body and merged first check jointly may shorten the initial ten-layer checkpoint. No shorter zero-cost construction was verified here.
\end{lstlisting}
\end{minipage}
\caption{Scratchpad notes from the Steane to Reed-Muller switch search of
Fig.~\ref{fig:recordratewordcount}, by (a) the base, (b) the strong and (c) the elite tier.
All three notes come from island 0 and accompany a new record. The record changes the
score (silent-cost, undecodable-cost, forfeit-cost, depth) of Eq.~\eqref{eq:score_gates}
from $(0,0,0,21)$ to $(0,0,0,20)$ in (a), from $(0,0,0,41)$ to $(0,0,0,38)$ in (b), and
from $(0,0,0,31)$ to $(0,0,0,26)$ in (c). The notes use the verifier's internal names:
\texttt{herald1} and \texttt{waste1}, or the herald and waste costs, which denote 
$\text{undecodable}_1$ and $\text{forfeited}_1$. ``discard costs'' means either of them.
Since $t=1$ here, these equal the undecodable-cost and forfeit-cost. Their lengths, 173, 206 and 445 words, are close to the means of their tiers, 169, 221
and 452 words, as shown in Fig.~\ref{fig:recordratewordcount}.}
\label{fig:scratchpad}
\end{figure*}

\FloatBarrier

\end{document}